\documentclass[aps,prl,twocolumn,superscriptaddress,longbibliography]{revtex4-2}
\usepackage{amsmath,amssymb,bm}
\usepackage{graphicx}
\usepackage{xcolor}
\usepackage[colorlinks=false,linkbordercolor=blue,citebordercolor=blue,urlbordercolor=blue,pdfborder={0 0 1}]{hyperref}
\hypersetup{hypertexnames=false} 
\usepackage{physics}
\usepackage{float}
\usepackage{placeins}

\renewcommand{\theHequation}{M\arabic{equation}}
\renewcommand{\theHfigure}{M\arabic{figure}}
\renewcommand{\theHtable}{M\arabic{table}}

\begin{document}

\title{Programmable Quantum Photonic Interfaces for Quantum Networking}

\author{Siavash Mirzaei-Ghormish}
\affiliation{Department of Electrical and Computer Engineering, Brigham Young University, Provo, Utah 84602, USA}
\author{Mahmoud Jalali Mehrabad}
\affiliation{Research Laboratory of Electronics, Massachusetts Institute of Technology, Cambridge, Massachusetts 02139, USA}
\author{Helaman Flores}
\affiliation{Research Laboratory of Electronics, Massachusetts Institute of Technology, Cambridge, Massachusetts 02139, USA}
\author{Dirk Englund}
\affiliation{Department of Electrical Engineering and Computer Science, Massachusetts Institute of Technology, Cambridge, Massachusetts 02139, USA}
\author{Ryan M. Camacho}
\affiliation{Department of Electrical and Computer Engineering, Brigham Young University, Provo, Utah 84602, USA}

\date{\today}

\begin{abstract}
Quantum networks require interfaces translating memory photons to telecom wavelengths while controlling spatial modes; tasks performed by separate components today. We present a programmable alternative: a structured pump writes a virtual Bragg grating enabling simultaneous spatio-spectral conversion and real-time controlling of emission. Using a LiNbO$_3$ whispering-gallery resonator, we demonstrate 93\% spatial coupling and bidirectional conversion between 736\,nm and 1347\,nm. This reconfigurable interface eliminates cascaded losses and hardware modifications.
\end{abstract}

\maketitle

\section{Introduction} 

Long-distance quantum communication requires photons emitted by stationary memories to be translated to the telecom band and routed efficiently~\cite{kimble2008quantum,afzelius2015quantum,zaske2012visibletelecom}. Standalone frequency converters span trapped ions, atoms, color centers, and quantum dots~\cite{bock2018high,krutyanskiy2019light,ikuta2018polarization,van2020long,luo2022postselected,schafer2025two,bernien2013heralded,bersin2024telecom,geus2024low,weber2019two,delteil2016generation,stockill2017phase}, yet spatial-mode engineering still depends on static gratings, tapers, or cavities~\cite{faraon2011resonant,riedrich2012one,hausmann2012integrated,albrecht2013coupling,mouradian2017rectangular,li2015efficient,ding2016ondemand}. Splitting the tasks across components adds loss, demands alignment, and offers no post-fabrication reconfiguration.
Existing integrated converters rely on lithographically fixed gratings or domain inversion~\cite{guo2016alnshg,cai2012integrated,fejer1992qpm}, and metasurface emitters shape far-field patterns without spectral translation~\cite{genevet2017recent,faraon2011resonant}, fixing the spectral and spatial response after fabrication. Related spatial-mode conversion has also been explored in waveguide arrays for entanglement generation~\cite{peruzzo2010quantumwalks}.

We introduce a theory for a single, programmable memory-photon interface that overcomes these constraints. A structured pump writes a virtual Bragg grating whose Fourier content is set entirely by the pump hologram, supplying the momentum and overlap needed for both spectral conversion and spatial routing in any multimode resonator. For clarity we instantiate the idea with a whispering-gallery geometry, where the harmonic content maps directly onto azimuthal indices, but the analysis extends to other cavities by projecting the pump onto their native mode bases. In every case the same optical drive simultaneously enforces the conversion selection rules, selects bright supermodes, and shapes the far-field emission. Because the grating is virtual (inscribed by light rather than etched into the device), the interface is bidirectional, alignment-free, and can be reprogrammed in situ as network requirements evolve.

\begin{figure}
    \centering
    \includegraphics[width=0.5\textwidth]{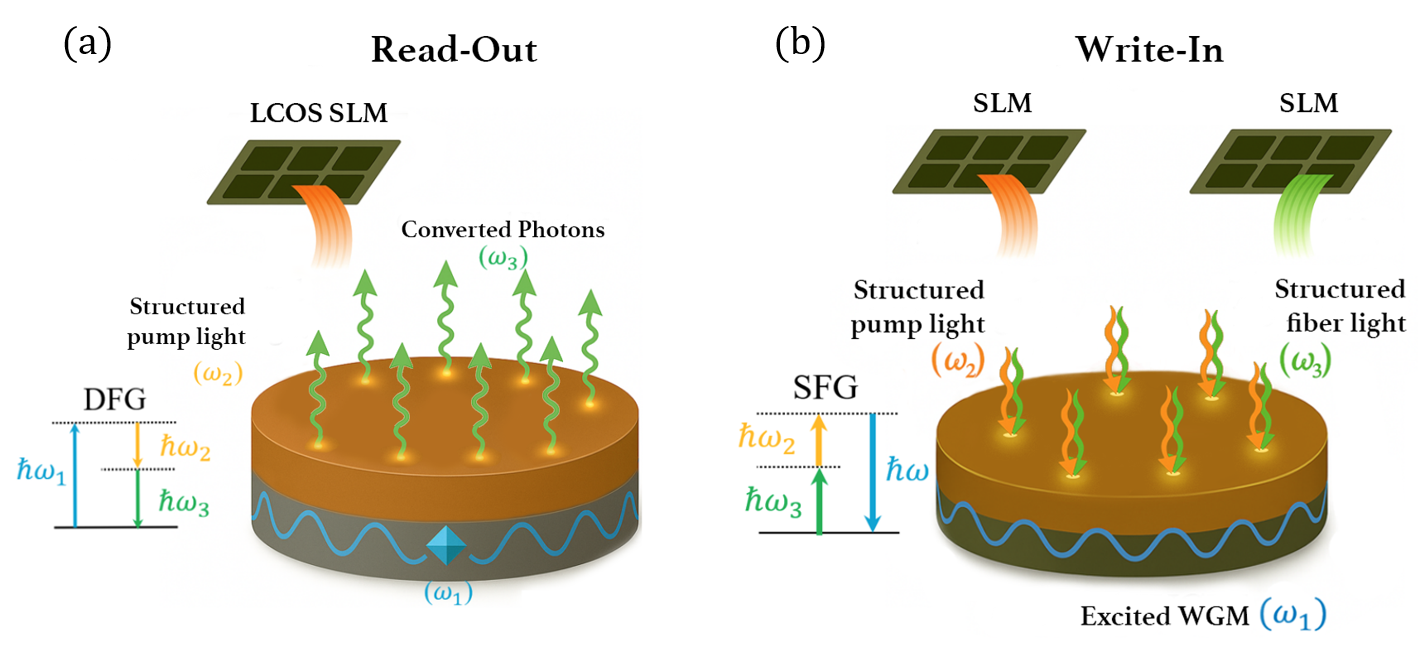}
    \caption{(a) Concept and device: a shaped, $N$-site pump writes a virtual Bragg grating on a $\chi^{(2)}$ ring co-axial with a diamond WGM hosting a SiV (parameters in Supplementary Table~2). (b) DFG (read-out) spectrum showing telecom emission at $\lambda_3=1347\,\mathrm{nm}$; dominant mode $(M,p,q)=(21,1,2)$ at 736\,nm. (c) SFG (write-in) spectrum around 738\,nm when a Gaussian fiber mode and the $N=23$ pump excite the cavity.}
    \label{figure1}
\end{figure}

To provide a concrete design, we specialize in a WGM resonator (Fig.~\ref{figure1}(a)), realized as a hybrid diamond disk with an embedded SiV center and a thin-film LiNbO$_3$ overlayer that supplies the second-order nonlinearity. The diamond layer is made sufficiently thin to enhance modal confinement within the nonlinear medium, thereby increasing interaction strength.

We first model the readout process where a photon emitted at frequency $\omega_1$ couples into a WGM mode $\hat{\mathbf{E}}_1(\mathbf{r},t)$ with azimuthal order $M$. A structured pump field $\mathbf{E}_2(\mathbf{r},t)$ drives DFG to generate a telecom photon $\hat{\mathbf{E}}_3(\mathbf{r},t)$ at $\omega_3 = \omega_1 - \omega_2$. The WGM is TE-like (polarized along $\hat{\boldsymbol{\rho}}$), while the pump is $\hat{\boldsymbol{\phi}}$-polarized. The resulting nonlinear polarization contains right-hand and left-hand side polarizations (RHCP and LHCP) components. When phase-matching conditions and radiative conditions are met, the telecom photon radiates into free space. Detailed derivations are provided in the Supplementary Information.

\begin{figure*}
    \centering
    \includegraphics[width=1\textwidth]{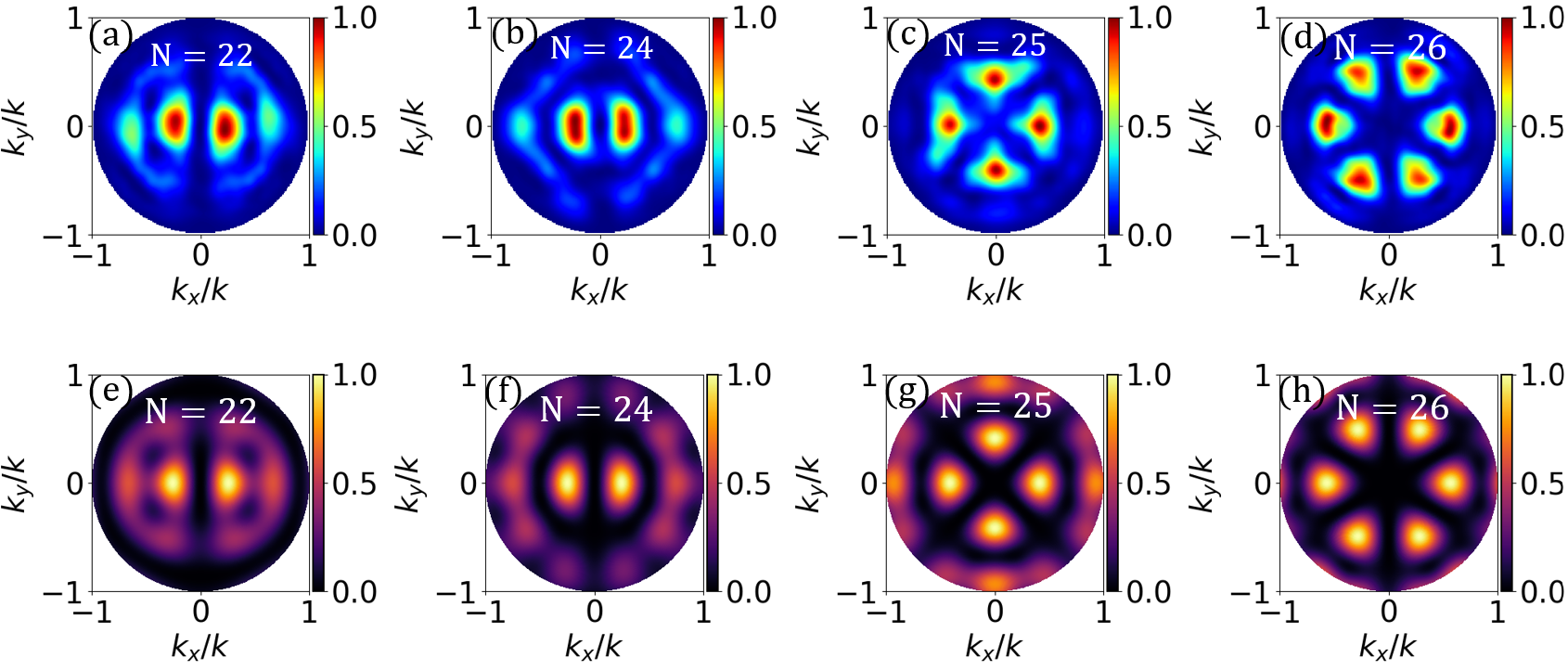}
\caption{Programmable far field. Each panel plots the normalised intensity $I(\theta,\phi)$ for $0\leq\theta\leq30^{\circ}$ (radial axis) and $0\leq\phi<2\pi$ (horizontal axis); color bars are normalised to the peak of each simulation. Varying the pump sampling $N$ at fixed $M=21$ (dominant harmonic $\ell=1$) changes the azimuthal periodicity $2|M-\ell N|$, switching between two-lobe patterns, on-axis brightening ($|M-\ell N|=2$), and multi-lobe morphologies, consistent with the selection rules and Eq.~\eqref{equation6}.}
    \label{figure2}
\end{figure*}

The pump field must carry azimuthal phase structure to satisfy angular momentum conservation. One could create this using $N$ discrete azimuthal sites at $\phi_n = 2\pi n/N$, yielding a Fourier series
$p(\phi)=\sum_{n=0}^{N-1}\delta(\phi-\phi_n)=\frac{N}{2\pi}\sum_{\ell\in\mathbb Z}e^{i\ell N\phi}$
containing harmonics at integer multiples of $N$. Alternatively, a smooth annular pump with helical phase $e^{im\phi}$ provides the same momentum matching. The angular matrix element
\(
\int_0^{2\pi}\!d\phi\;e^{i\Delta m\,\phi}\,p(\phi)
\)
is nonzero only when $\Delta m$ matches an available harmonic: $\Delta m=\ell N$ for the discrete case, or $\Delta m = m$ for the smooth case.
Independently, the basis geometry of the chosen in-plane polarizations
(TE WGM $\hat{\boldsymbol\rho}$ and pump $\hat{\boldsymbol\phi}$) inserts a fixed
$2\phi$ harmonic:
$\hat{\boldsymbol\rho}\otimes\hat{\boldsymbol\phi}\propto
-\cos 2\phi\,\hat{\mathbf x}+\sin 2\phi\,\hat{\mathbf y}$.
After contraction with $\overline{\overline{\chi}}^{(2)}$ this contributes a fixed
offset $\Delta m_{\text{geom}}=\pm 2$. Other pump configurations yield different rules. Combining these two facts with the WGM
phase $e^{\pm iM\phi}$  and applying radiation conditions, gives the general azimuthal selection rules:.

\begin{align}
\ell_{+} &= -\ell'_{-} = M - N + 2,  \quad
\ell'_{+} &= -\ell_{-} = M - N - 2,
\label{equation1}
\end{align}

where $\hbar\ell_{\pm}$ are OAMs of RHCP components, and $\hbar\ell'_{\pm}$ of LHCP.  Radiation requires $|\ell_i|/R<k_3$, which selects a small set of diffraction orders
$\ell$ (often a single order per direction). For our parameters, the radiative
window yields $\ell=-1$ (CW) and $\ell=+1$ (CCW),

An equivalent geometric condition for radiative mode \(i\) should be $|R| > |\ell_i|/k_3$. A full tensor-based derivation, including the explicit contraction yielding $d_{R,L}$, is provided in \hyperref[supp:secI]{Supplementary Sec.~I}. There we show that the quasi-phase-matching harmonics normally supplied by etched gratings can instead be written optically by the pump, allowing the selection rules to be reprogrammed without modifying the resonator (Fig.~\ref{figure2}). Although every integer $\ell$ contributes in principle, our simulations confirm that $|\ell|=1$ dominates for the device parameters in Fig.~\ref{figure1}; we suppress explicit $\ell$ dependence in what follows unless stated otherwise.

These rules resemble those in Bragg gratings~\cite{cai2012integrated,fejer1992qpm} but stem from nonlinear symmetry, not material periodicity. The azimuthal modulation of the pump thus acts as a virtual Bragg grating, whose Fourier components mediate specific OAM transitions. Unlike etched gratings or poled domains, our approach writes the phase-matching structure optically, enabling in-situ reconfiguration without device modification. By dynamically controlling the pump's complex amplitude and polarization, one can selectively engage different radiative channels, enabling tunable emission directionality and angular momentum transfer beyond static grating capabilities.

Within the rotating-wave approximation (RWA), we work in the interaction picture with respect to the free Hamiltonian so that rapidly oscillating, nonresonant terms are discarded. The DFG Hamiltonian is:

\begin{align}
\hat H_{\mathrm{DFG}} &= 
\hbar \left|g_{\mathrm{eff}}^{(+)}\right|\left(
\hat b_{+}^{\dagger}\hat a_{1+}
+
\hat a_{1+}^{\dagger}\hat b_{+}
\right) 
\nonumber \\
&\quad +
\hbar \left|g_{\mathrm{eff}}^{(-)}\right|\left(
\hat b_{-}^{\dagger}\hat a_{1-}
+
\hat a_{1-}^{\dagger}\hat b_{-}
\right),
\label{equation2}
\end{align}

where $\hat a_{1\pm}$ denote CW and CCW WGMs, and $\hat b_{\pm}$ the bright-mode superpositions. The telecom output naturally decomposes into four channels (CW/CCW propagation with R/L circular polarization), which we combine into two bright supermodes. The microscopic overlaps $g_{R,L}^{(\pm)}$ can in general be complex. Writing $g_{\mathrm{eff}}^{(\pm)} = |g_{\mathrm{eff}}^{(\pm)}| e^{i\theta_{\pm}}$ and redefining $\hat a_{1\pm}\rightarrow e^{i\theta_{\pm}}\hat a_{1\pm}$ absorbs the global phase into the WGM operators, so Eq.~\eqref{equation2} has the beam-splitter form with a strictly real coupling rate. We adopt the same phase convention when assembling the bright supermodes:

\begin{align}
\hat b_{+} &= 
\frac{g^{(+)}_{R}\hat a^{(\mathrm{cw})}_{3,R}
+g^{(+)}_{L}\hat a^{(\mathrm{cw})}_{3,L}}
{g_{\mathrm{eff}}^{(+)}}, \\
\hat b_{-} &=
\frac{g^{(-)}_{R}\hat a^{(\mathrm{ccw})}_{3,R}
+g^{(-)}_{L}\hat a^{(\mathrm{ccw})}_{3,L}}
{g_{\mathrm{eff}}^{(-)}},
\label{equation3}
\end{align}

where $\hat a_{3,R/L}^{(\mathrm{cw})}$ annihilate right/left-hand circularly polarized photons propagating clockwise in the radiative continuum, and $\hat a_{3,R/L}^{(\mathrm{ccw})}$ the analogous counterclockwise operators; they obey canonical bosonic commutators, e.g.\ $[\hat a_{3,R}^{(\mathrm{cw})},\hat a_{3,R}^{(\mathrm{cw})\dagger}]=1$. The bright modes $\hat b_{\pm}$ therefore inherit the same algebra and act as normalized superpositions of the four output channels. With this notation $g_{\mathrm{eff}}^{(\pm)} = \sqrt{|g_{R}^{(\pm)}|^{2} + |g_{L}^{(\pm)}|^{2}}$, ensuring $[\hat b_{\pm},\hat b_{\pm}^{\dagger}] = 1$ and $[\hat b_{+},\hat b_{-}^{\dagger}]=0$. The detailed assembly of this bright-mode Hamiltonian, including its reuse in the SFG write-in problem, appears in \hyperref[supp:secI]{Supplementary Sec.~I}.

The DFG conversion efficiency, defined as the fractional photon transfer from WGM to bright mode, follows from beam-splitter dynamics:

\begin{equation}
\eta^{\pm}_{\mathrm{DFG}} =
\frac{\langle \hat b_{\pm}^\dagger(T) \hat b_{\pm}(T) \rangle}
{\langle \hat a_{1,\pm}^\dagger(0) \hat a_{1,\pm}(0) \rangle}
= \sin^2\big( g^{\pm}_{\mathrm{eff}} T \big)
= \sin^2\big( \gamma^{\pm}_{\phi} L_{\phi} \big),
\label{equation4}
\end{equation}

where \( \gamma^{\pm}_{\phi} = g^{\pm}_{\mathrm{eff}} / v_{\phi} \) is the azimuthal nonlinear coupling rate, and \( L_{\phi} = v_{\phi} T \) is the effective azimuthal interaction length. Here, \( v_{\phi} \) denotes the group velocity of the WGM mode in the azimuthal direction. Equation~\eqref{equation4} therefore describes a single-pass interaction that neglects pump depletion and cavity recirculation; cavity-enhanced or high-gain operation would replace the sinusoidal response with the familiar $\tanh^2$ saturation.

The mean photon number in a specific radiative output channel, such as the right-hand circularly polarized component in the CW manifold, is given by

\begin{equation}
n_{R\pm} = 
\langle \hat a^{\dagger}_{3, R\pm} \hat a_{3, R\pm} \rangle 
= \frac{ |g^{(\pm)}_{R}|^{2} }{ (g^{(\pm)}_{\mathrm{eff}})^2 }
\, \eta^{\pm}_{\mathrm{DFG}} \, 
\langle \hat a^{\dagger}_{1\pm}(0) \hat a_{1\pm}(0) \rangle.
\label{equation5}
\end{equation}

Similar expressions hold for the remaining three radiative channels. 

To evaluate the far-field intensity, we model the nonlinear interaction as arising from a discrete array of subwavelength emitters. Each azimuthal site acts as an effective dipole radiating a photon at frequency \(\omega_3\), with the total field resulting from coherent superposition across all interaction points. Four distinct dipole sets correspond to the four radiative channels permitted by the angular momentum selection rules [Fig.~\ref{figure1}(b)]. Collecting the radial and axial Green-function factors we write \(|\mathcal N g|^2 = \big(\tfrac{k_3^2 \chi_{\mathrm{eff}} L_z}{4\pi \varepsilon_0 r}\big)^2 |g(\theta,r)|^2\), where the explicit envelope \(g(\theta,r)\) is given in Supplementary Eqs.~(20) to (23). The far-field intensity \(I(\theta,\phi)\) is then

\begin{align}
I(\theta,\phi)
&= |\mathcal N g|^{2}\Big[
n_{R+}\,J^2_{|\ell_{+}|}(x)+n_{R-}\,J^2_{|\ell_{-}|}(x)
\notag\\[-2pt]
&\qquad
+2\,|\Gamma_{R}|\,J_{|\ell_{+}|}(x)J_{|\ell_{-}|}(x)\,
\cos\!\big( \Delta m_R\,\phi+\psi_R \big)
\Big]
\notag\\
&\quad+|\mathcal N g|^{2}\Big[
n_{L+}\,J^2_{|\ell'_{+}|}(x)+n_{L-}\,J^2_{|\ell'_{-}|}(x)
\notag\\[-2pt]
&\qquad
+2\,|\Gamma_{L}|\,J_{|\ell'_{+}|}(x)J_{|\ell'_{-}|}(x)\,
\cos\!\big( \Delta m_L\,\phi+\psi_L \big)
\Big],
\label{equation6}
\end{align}

where \(x = k_3 R \sin\theta\), with \(k_3\) the wavevector of the emitted photon, and \(R\) the resonator radius. The quantities \(n_{R\pm} = \langle \hat a^{\dagger}_{3,R\pm} \hat a_{3,R\pm} \rangle\) and \(n_{L\pm} = \langle \hat a^{\dagger}_{3,L\pm} \hat a_{3,L\pm} \rangle\) denote the mean photon numbers in the respective channels. Coherence terms are defined as \(\Gamma_{R} = \langle \hat a^{\dagger}_{3,R+} \hat a_{3,R-} \rangle = |\Gamma_R| e^{i\psi_R}\), and similarly for \(\Gamma_L\). The definitions of $\Delta m_{R,L}$ and the associated coherence matrices are summarised in \hyperref[supp:secII]{Supplementary Sec.~II}; Eq.~\eqref{equation6} is the expression evaluated in the results below.

The angular structure of the intensity pattern is governed by the spatial harmonic order \(\Delta m_R=\Delta m_L=2|M - \ell N|\), which sets the number of interference lobes in \(\phi\). In particular, when \(|M - \ell N| = 2\), the radiation exhibits constructive interference along the optical axis (\(\theta = 0\)), resulting in strong on-axis emission and enhanced fiber coupling efficiency. A full derivation of Eq.~\eqref{equation6} is provided in the Supplementary Information.

\begin{figure*}
    \centering
    \includegraphics[width=1\textwidth]{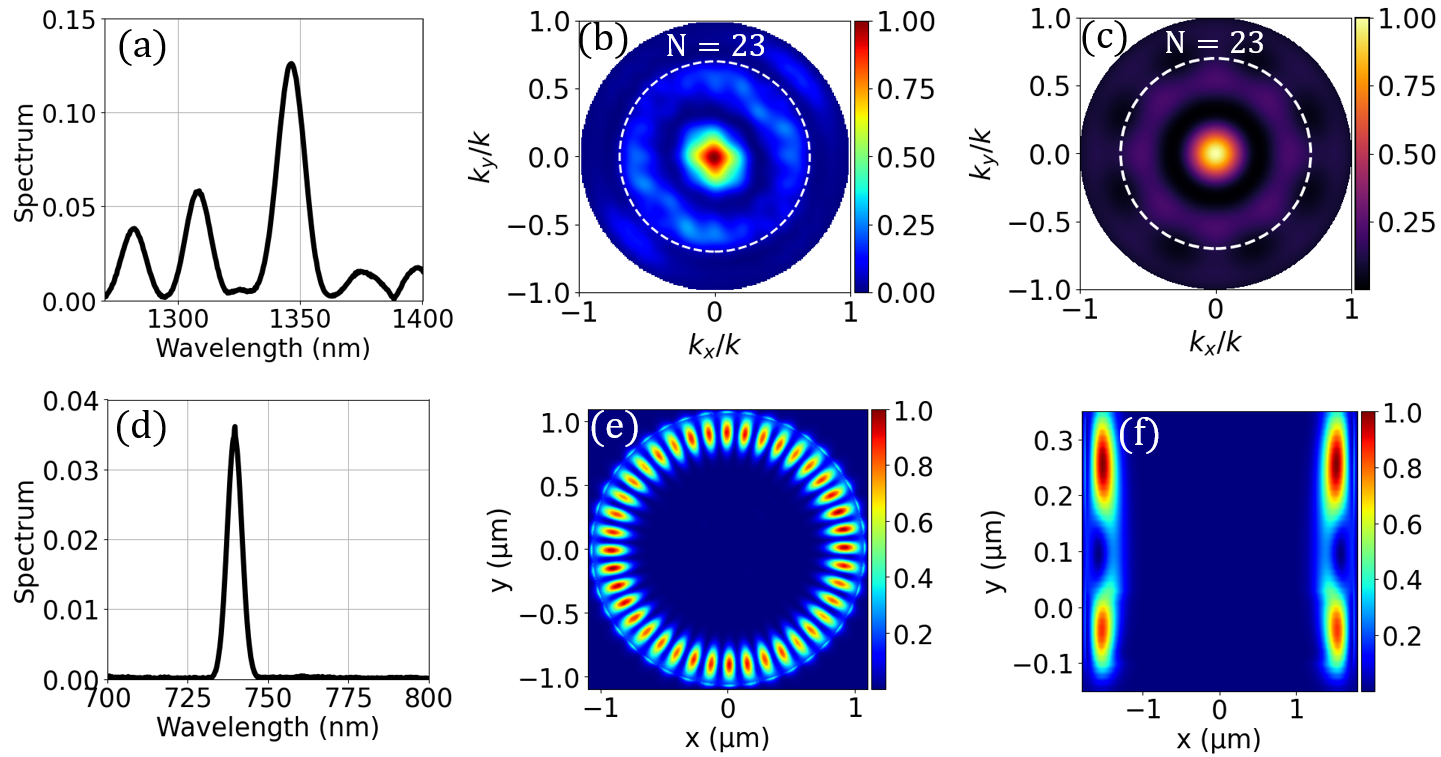}
    \caption{(a) DFG readout spectrum: converted telecom peak at $\lambda_3=1347$\,nm driven from the $(M,p,q)=(21,1,2)$ WGM. (b) Far-field intensity for $N=23$ from FDTD; (c) analytic bright-mode prediction for the same case. (d) SFG write-in spectrum around $\lambda_1\approx736$\,nm under the same pump. (e) Spatial mode of the coupled WGM (normalized intensity). (f) Bright-mode population versus azimuthal interaction length for the SFG process; color scales in panels (b) to (e) are normalized to their respective maxima.}
    \label{figure3}
\end{figure*}

The virtual grating picture is fully reciprocal. In the write-in process, a telecom-band fiber photon at frequency \(\omega_3\) interacts with a strong, undepleted pump at \(\omega_2\) to upconvert into a WGM mode near \(\omega_1 = \omega_2 + \omega_3\) via sum-frequency generation (SFG). The generated frequency must lie within the cavity linewidth of the target WGM, such that \( |\omega_1 - \omega_{M,p,q}| \le \kappa/2 \), where \( (M, p, q) \) denote the azimuthal, radial, and longitudinal mode indices, and \(\kappa\) is the loaded linewidth.

To maximize nonlinear overlap, both the pump and fiber fields are shaped with an identical \(N\)-fold azimuthal sampling. With the fiber mode polarized along \(\hat{\boldsymbol\rho}\) and the pump along \(\hat{\boldsymbol\phi}\), the nonlinear polarization radiates both CW and CCW WGM modes. Enforcing angular phase-matching and the guiding condition yields the same lattice rule as in the read-out process:
\begin{equation}
M = |\ell N \pm 2|,\qquad \ell\in\mathbb Z,
\label{equation7}
\end{equation}
where the sign corresponds to RHCP/LHCP emission. As in the DFG case, the \(N\)-site sampling writes a virtual Bragg grating, supplying orbital angular momentum in multiples of \(N\), while the \(\pm 2\) shift arises from the polarization configuration. For the parameters in Fig.~\ref{figure3}, the $\ell=\pm1$ channels dominate, so the leading WGMs have $M=N\pm2$.

The SFG interaction Hamiltonian under RWA takes the form:

\begin{equation}
\hat H_{\text{SFG}} 
= \hbar \left|G_{\mathrm{eff}}\right| \left( \hat b^\dagger \hat a_3 + \hat a_3^\dagger \hat b \right).
\label{equation8}
\end{equation}

where \(\hat a_3\) is the annihilation operator of the incoming fiber photon, and \(\hat{b}\) is a bright supermode defined as

\begin{equation}
\hat b = \frac{1}{G_{\mathrm{eff}}} \sum_i G_i \hat a_{1,i}, 
\quad 
G_{\mathrm{eff}} = \left( \sum_i |G_i|^2 \right)^{1/2},
\label{equation9}
\end{equation}

\(\hat a_{1,i}\) denotes the \(i^{\text{th}}\) WGM channel coupled by the pump; the \(G_{\mathrm{eff}}\) is the effective coupling coefficient. Microscopic overlaps obey $G_i = |G_i| e^{i\varphi_i}$, and redefining $\hat a_{1,i}\rightarrow e^{i\varphi_i}\hat a_{1,i}$ renders $G_{\mathrm{eff}}$ real and positive, matching the beam-splitter form of Eq.~\eqref{equation8}.

The bright supermode satisfies canonical bosonic commutation relations \([\hat b, \hat b^\dagger] = 1\). 

Because the $N$-site pump only excites the \(M=N\pm2\) azimuthal manifolds from Eq.~\eqref{equation7}, the bright mode \(\hat b\) is precisely the coherent sum of these channels, so Eq.~\eqref{equation8} reduces to the single beam-splitter Hamiltonian with coupling rate \(G_{\mathrm{eff}}\) quoted above.

The corresponding upconversion efficiency into the bright WGM supermode after interaction time \(T\) (or azimuthal propagation length \(L_\phi = 2\pi R\)) is given by
\begin{equation}
\eta_{\mathrm{SFG}} 
= \frac{\langle \hat b^\dagger(T) \hat b(T) \rangle}{\langle \hat a_3^\dagger(0) \hat a_3(0) \rangle}
= \sin^2\!\big( G_{\mathrm{eff}} T \big)
= \sin^2\!\big( \gamma_\phi L_\phi \big),
\label{equation10}
\end{equation}
where \(\gamma_\phi = G_{\mathrm{eff}} / v_\phi\) is the effective azimuthal nonlinear coupling rate, and \(v_\phi = L_\phi / T\) is the group velocity along the ring.

\section{Numerical Validation}

To verify the analytical model, we performed classical finite-difference time-domain (FDTD) simulations using Lumerical. Because the released solver supports only isotropic second-order susceptibilities, we emulate LiNbO$_3$ with a scalar coefficient equal to the root-mean-square of the anisotropic combinations, setting $\chi_{11} = \chi_{22} = 4.5 \times 10^{-11}\,\mathrm{m/V}$ (denoted $\chi_{\mathrm{iso}}$) and all other tensor elements to zero. The simulated conversion amplitudes are then rescaled by $d_{R}/\chi_{\mathrm{iso}}$ and $d_{L}/\chi_{\mathrm{iso}}$ when compared with the analytic couplings, restoring the correct RHCP/LHCP weighting despite the isotropic surrogate (see Supplement Sec.~V).

For the out-coupling process, the simulated geometry consists of a diamond ring resonator of radius \(1.6\,\mu\mathrm{m}\), radial width \(200\,\mathrm{nm}\), and thickness \(100\,\mathrm{nm}\), overlaid with a \(280\,\mathrm{nm}\) LiNbO\(_3\) layer. To cleanly demonstrate the physics of momentum matching and selection rules, the pump field is modeled as \(N=23\) discrete Gaussian beams (waist \(a = 120\,\mathrm{nm}\), wavelength \(\lambda_p = 1623\,\mathrm{nm}\)) positioned at \(\phi_n=2\pi n/N\) with appropriate phase. This sub-diffraction waist is used only as an idealized proxy to isolate the selection-rule physics; the experimentally feasible smooth-annulus implementation (waist \(\sim0.6\,\mu\)m limited by NA) produces the same azimuthal order \(m=23\) and far-field structure. The NA-limited pump layout, steering span, and design example are detailed in Supplement Sec.~III, which also estimates \(\sim8\times10^{4}\) addressable sites on a 10\,$\mu$m grid for N$\times$1 switching.

Within this surrogate the spatial symmetries generate the $\pm2$ azimuthal factors dictated by Eq.~\eqref{equation1}, and the RMS rescaling maps them onto the anisotropic coefficients $d_{R}$ and $d_{L}$. We verify consistency by noting that the azimuthal phase factor for the \((M=21,\lambda_{1}=736\,\mathrm{nm})\) mode and \(R=1.6\,\mu\mathrm{m}\) implies \(n_{\mathrm{eff}}\approx M\lambda_{1}/(2\pi R)\approx1.54\), and the DFG relation \(1/\lambda_{3}=1/\lambda_{1}-1/\lambda_{2}\) with \(\lambda_{2}=1623\,\mathrm{nm}\) gives \(\lambda_{3}\approx1346.9\,\mathrm{nm}\), matching the simulated telecom output. In the far field, the intensity pattern simplifies to:

\begin{align}
I(x, \phi)
&= 2A \Big[
J^2_{|\ell_+|}(x) + J^2_{|\ell_+'|}(x)
+ J^2_{|\ell_-|}(x) + J^2_{|\ell_-'|}(x)
\notag\\
&\qquad
+ 2 J^2_{|\ell_+|}(x) \cos\!\big(2 \ell_+ \phi\big)
+ 2 J^2_{|\ell_-|}(x) \cos\!\big(2 \ell_- \phi\big)
\Big],
\label{equation12}
\end{align}
using $J_{-m}(x)=(-1)^m J_m(x)$ so intensities depend on $|\ell|$. For $N > M$, the dominant low-NA modulation arises from $\cos\!\big(2 \ell_+ \phi\big)$ with $2\ell_+=2(M-N+2)$; for $N < M$, from $\cos\!\big(2 \ell_- \phi\big)$ with $2\ell_-=2(M-N-2)$. When $|M-N|=2$, the low-NA intensity is maximized on-axis.

Figure~\ref{figure3}(a) shows the simulated readout spectrum, revealing WGM modes at \(\lambda = 730\,\mathrm{nm}, 736\,\mathrm{nm}, 747\,\mathrm{nm},\) and \(751\,\mathrm{nm}\). The TE-like mode at \(\lambda = 736\,\mathrm{nm}\), identified as \((M, p, q) = (21, 1, 2)\), shows the strongest nonlinear response and yields converted emission at \(\lambda_3 = 1347\,\mathrm{nm}\).

Figure~\ref{figure2} compares these analytical predictions ((e)-(h)) with full-wave simulations ((a)-(d)) for representative pump samplings. It toggles between multi-lobe and on-axis emission, delivering \(\eta_{\mathrm{spatial}}\approx93\%\) in the $N=23$  case (Figure \ref{figure3}(b),(c)) (full efficiency breakdown in \hyperref[supp:secIII]{Supplementary Sec.~III}). Figures~\ref{figure3} (d)-(f) summarize the reciprocal SFG process using the same hologram; the associated spectra, mode profiles, and pump-induced shifts appear in \hyperref[supp:secI]{Supplementary Sec.~I}.

Experimentally feasible holograms are generated with standard near-infrared LCOS SLMs; required pump phase stability ($<1^{\circ}$) and spatial positioning ($<0.5\,\mu$m) are compatible with commercial hardware and detailed in \hyperref[supp:secIII]{Supplementary Sec.~III} along with optical layout and pump-power scalings. \hyperref[supp:secIV]{Supplementary Sec.~IV} quantifies noise processes (Raman, leakage, thermal) and finds them well below the conversion rates quoted here.

These simulations confirm that a structured pump can write a virtual Bragg grating that mediates simultaneous spectral and spatial photon conversion. Once the pump is projected onto any cavity's eigenmodes the same hologram fixes the selection rules, bright-mode structure, and far-field routing, turning spectral and spatial reconfiguration into a software task. We have established that virtual Bragg gratings enable programmable quantum interfaces by unifying frequency conversion and spatial mode control in a single reconfigurable element. This integrated approach eliminates separate mode-matching components, reducing cascade complexity while maintaining comparable efficiency and enabling in-situ reconfiguration to match different emitters or network requirements without hardware modification. All required experimental ingredients already exist: $Q>10^{6}$ resonators~\cite{guo2016alnshg}, stable SLM holography, and efficient frequency converters.  Next steps include validating the pump-programmed far fields with classical probes, demonstrating cavity-enhanced DFG/SFG with heralded photons, and adapting holograms in repeater nodes to match remote emitters, establishing practical pathways toward scalable quantum networking.

\clearpage
\onecolumngrid
\appendix*
\setcounter{section}{0}
\setcounter{equation}{0}
\setcounter{figure}{0}
\setcounter{table}{0}
\renewcommand{\theequation}{S\arabic{equation}}
\renewcommand{\thefigure}{S\arabic{figure}}
\renewcommand{\thetable}{S\arabic{table}}
\renewcommand{\theHequation}{S\arabic{equation}}
\renewcommand{\theHfigure}{S\arabic{figure}}
\renewcommand{\theHtable}{S\arabic{table}}

\section*{Supplementary Information for:\\
Programmable Quantum Photonic Interfaces for Quantum Networking}

This Supplementary Information provides complete theoretical derivations and experimental implementation details for the virtual Bragg grating approach described in the main text. The material is organized as follows:

\textbf{Section~I (Quantum Frequency Conversion):} Derives the selection rules $M = \ell N \pm 2$ and bright-mode Hamiltonian from full $\chi^{(2)}$ tensor analysis. Treats both DFG (readout) and SFG (write-in) processes, providing the $d_{R,L}$ coupling coefficients and conversion efficiency expressions referenced in main text Eqs.~(1) to (4) and (8) to (10).

\textbf{Section~II (Far-Field Emission):} Connects the nonlinear polarization to observable far-field patterns. Derives the intensity expression Eq.~(6) from dipole-array Green functions, defines coherence parameters $\Delta m_{R,L}$, $\Gamma_{R,L}$, and provides operator normalization conventions for photon number calculations.

\textbf{Section~III (Pump Implementation):} Details the experimental realization using LCOS spatial light modulators. Subsection~A lays out bandwidth and selection-rule constraints; Subsection~B describes optical layout and calibration; Subsection~C gives design examples; Subsection~D estimates addressable array size for an N$\times$1 switch; Subsection~E derives pump-power scaling; Subsection~F quantifies reconfiguration bandwidth.

\textbf{Section~IV (Noise and Decoherence):} Quantifies Raman scattering, pump leakage, and thermal noise contributions, confirming all remain below conversion rates at the design operating point.

\textbf{Section~V (Numerical Verification):} Reports FDTD simulation methods, isotropic-surrogate rescaling procedure, and device parameters (Table~2). Validates analytical predictions for spectra (Fig.~1), far-field patterns (Fig.~2), and write-in dynamics (Fig.~3).


\section*{I. Quantum frequency conversion}
\phantomsection\label{supp:secI}

\textbf{Notation and conventions.} We adopt the following notation throughout: $\omega_1$ (visible SiV frequency), $\omega_2$ (pump frequency), $\omega_3$ (telecom output frequency) satisfy $\omega_3 = \omega_1 - \omega_2$ for DFG and $\omega_1 = \omega_2 + \omega_3$ for SFG. The WGM azimuthal quantum number is $M$, and the pump has $N$ azimuthal sites. Thus, the pump adds angular momentum of \(\ell N\), where $\ell \in \mathbb{Z}$ labels harmonics of the pump Fourier decomposition; throughout most derivations $|\ell|=1$ dominates. Moreover, polarization geometry shifts the total angular momentum by \(\pm 2\). The second-order susceptibility tensor $\overline{\overline{\chi}}^{(2)}$  contracts to effective RHCP/LHCP coupling coefficients  $d_{R,L}$ defined in Eq.~(9). Operator $\hat{a}_{1\pm}$ denotes CW/CCW WGM modes, $\hat{a}_{3,R/L}^{(\mathrm{cw/ccw})}$ labels radiative channels by helicity and propagation direction, and $\hat{b}_{\pm}$ are bright supermodes. Polarization basis vectors: $\ket{R} = (\hat{e}_x + i\hat{e}_y)/\sqrt{2}$ (RHCP), $\ket{L} = (\hat{e}_x - i\hat{e}_y)/\sqrt{2}$ (LHCP), with $\hat{\boldsymbol{\rho}} = \cos\phi\,\hat{e}_x + \sin\phi\,\hat{e}_y$ (radial) and $\hat{\boldsymbol{\phi}} = -\sin\phi\,\hat{e}_x + \cos\phi\,\hat{e}_y$ (azimuthal). All fields use $e^{-i\omega t}$ time dependence. Table~\ref{tab:operators} summarizes radiative operators.

\subsection*{A. Read-out process}

In the read-out direction a single SiV photon is converted to the telecom band through difference-frequency generation (DFG) with the programmable pump. The derivation proceeds in three stages: we first specify the quantized cavity and radiative fields, then evaluate the structured pump together with the induced nonlinear polarization, and finally project the source onto the allowed angular-momentum channels that enter the effective Hamiltonian. Throughout this subsection the pump is treated as an undepleted classical drive, whereas the SiV-coupled WGM and the telecom output are quantized.

\paragraph*{Quantized fields.}
The SiV emission couples to a WGM with azimuthal number $M$, frequency $\omega_1$, and radial wavevector $k_1$. The electric field operator is
\begin{equation}
\hat{\mathbf{E}}_1(\mathbf{r}, t) =
\Big[
\hat{a}_{1+}(\phi)\, J_M(k_1 \rho)\, e^{+iM\phi}
+
\hat{a}_{1-}(\phi)\, J_M(k_1 \rho)\, e^{-iM\phi}
\Big]
e^{\pm i k_{1z} z}\, e^{-i\omega_1 t}\, \hat{\boldsymbol{\rho}}
\;+\; \text{H.c.}
\label{supp:eq1}
\end{equation}
Here $\hat{a}_{1\pm}(\phi)$ are slowly varying envelope operators for clockwise (CW) and counter-clockwise (CCW) WGMs, $J_M(k_1\rho)$ is the radial Bessel profile, and $\hat{\boldsymbol{\rho}}$ denotes the TE-like polarization.

\paragraph*{Structured pump and nonlinear polarization.}
The undepleted pump at $\omega_2$ is incident from $-z$ and azimuthally structured at $N$ discrete angles $\phi_n=2\pi n/N$:

\begin{equation}
\mathbf{E}_2(\mathbf{r}, t) =
\alpha\, J_P(k_2 \rho)\,
\sum_{n=0}^{N-1}
\delta\!\Big( \phi - \tfrac{2\pi n}{N} \Big)\,
e^{-i k_{2z} z} e^{-i\omega_2 t}\, \hat{\boldsymbol{\phi}}
\;+\; \text{c.c.}
\label{supp:eq2}
\end{equation}
The discrete angular modulation has a Fourier expansion
\(
p(\phi) =
\sum_{n=0}^{N-1}\delta\!\Big(\phi-\tfrac{2\pi n}{N}\Big)
=
\frac{N}{2\pi}\sum_{\ell\in\mathbb{Z}} e^{i\ell N\phi}.\)

The second-order polarization that sources the telecom field at $\omega_3=\omega_1-\omega_2$ is

\begin{equation}
\hat{\mathbf{P}}^{(2)}(\mathbf{r}, t)
= \varepsilon_0\, \overline{\overline{\chi}}^{(2)} : \hat{\mathbf{E}}_1(\mathbf{r}, t)\, \mathbf{E}_2^{\,*}(\mathbf{r}, t).
\label{supp:eq3}
\end{equation}
Evaluating with Eqs.~(1) to (3) and the chosen polarizations ($\hat{\boldsymbol{\rho}}$ for the WGM and $\hat{\boldsymbol{\phi}}$ for the pump) gives

\begin{equation}
\hat{\mathbf{P}}^{(2)}(\mathbf{r}, t) =
\sum_{n=0}^{N-1}
\overline{\overline{\chi}}^{(2)} : \hat{\boldsymbol{\rho}}_n \hat{\boldsymbol{\phi}}_n\,
\delta(\phi-\phi_n)
\Big(
\hat{\eta}_+\,e^{iM\phi_n} + \hat{\eta}_-\,e^{-iM\phi_n}
\Big)
\,e^{(\pm i k_{1z} z+i k_{2z})}\, e^{-i\omega_3 t}
\;+\; \text{H.c.},
\label{eq:dipole_local}
\end{equation}
with operator coefficients
\(
\hat{\eta}_\pm = \varepsilon_0\, J_M(k_1 \rho)\, J_P(k_2 \rho)\, \hat{a}_{1\pm}(\phi)\, \alpha^{*}.
\)

Projecting into the $(\hat{\mathbf{x}},\hat{\mathbf{y}})$ basis at each site $\phi_n$ yields
\begin{equation}
\begin{aligned}
\hat{\mathbf{P}}_n
&=
\overline{\overline{\chi}}^{(2)} :
\Big(
-\frac{1}{2}\sin 2\phi_n\, \hat{\mathbf{x}}\hat{\mathbf{x}}
+ \frac{1}{2}\sin 2\phi_n\, \hat{\mathbf{y}}\hat{\mathbf{y}}\\
&\qquad\qquad+
\cos^2\!\phi_n\, \hat{\mathbf{x}}\hat{\mathbf{y}}
- \sin^2\!\phi_n\, \hat{\mathbf{y}}\hat{\mathbf{x}}
\Big)
\Big(
\hat{\eta}_+ e^{iM\phi_n} + \hat{\eta}_- e^{-iM\phi_n}
\Big)
e^{(\pm i k_{1z} z+i k_{2z})}\, e^{-i\omega_3 t}
+ \text{H.c.}
\end{aligned}
\label{eq:farfield_single}
\end{equation}

\noindent
We take the nonlinear medium to be $z$-cut LiNbO$_3$, whose contracted tensor has the non-vanishing entries $d_{31}=d_{32}$, $d_{33}$, $d_{15}=-d_{24}$, and $d_{22}=-d_{16}$. Rotating this tensor from the crystal frame into the local cylindrical basis gives in-plane coefficients
\begin{equation}
\mathcal{C}_x(\phi_n) = -d_{22}\cos 2\phi_n + d_{31}\sin 2\phi_n,
\qquad
\mathcal{C}_y(\phi_n) = d_{22}\sin 2\phi_n + d_{31}\cos 2\phi_n,
\label{eq:lnbo_coeffs}
\end{equation}
for the $\rho$-polarized WGM interacting with a $\phi$-polarized pump; contributions from $d_{33}$ and $d_{15}$ vanish for this polarization choice. The local nonlinear polarization therefore reads
\begin{equation}
\hat{\mathbf{P}}_n =
\Big(\hat{\eta}_+ e^{iM\phi_n} + \hat{\eta}_- e^{-iM\phi_n}\Big)
e^{(\pm i k_{1z} z+i k_{2z})}\, e^{-i\omega_3 t}\,
\big[
\mathcal{C}_x(\phi_n)\,\hat{\mathbf{x}} + \mathcal{C}_y(\phi_n)\,\hat{\mathbf{y}}
\big]
\;+\; \text{H.c.}
\label{supp:eq4}
\end{equation}

\paragraph*{Angular-momentum decomposition.}
After summing over all pump sites via the Poisson series, the total polarization operator becomes
\begin{equation}
\begin{aligned}
\hat{\mathbf{P}}^{(2)}
&= -\frac{N}{4\pi}
\sum_{\ell\in\mathbb{Z}}
\Big\{
\hat{\eta}_{+}
\Big[
d_{R}\,e^{i(\ell N+M+2)\phi}\ket{R}
+ d_{L}\,e^{i(\ell N+M-2)\phi}\ket{L}
\Big]\\
&\qquad\qquad+
\hat{\eta}_{-}
\Big[
d_{R}\,e^{i(\ell N-M+2)\phi}\ket{R}
+ d_{L}\,e^{i(\ell N-M-2)\phi}\ket{L}
\Big]
\Big\}
e^{(\pm i k_{1z} z+i k_{2z})}\, e^{-i\omega_3 t}
+ \text{H.c.}
\end{aligned}
\label{eq:farfield_intensity}
\end{equation}
where $\ket{R}=\tfrac{1}{\sqrt{2}}(\hat{\mathbf{x}}+i\hat{\mathbf{y}})$ and $\ket{L}=\tfrac{1}{\sqrt{2}}(\hat{\mathbf{x}}-i\hat{\mathbf{y}})$ denote right- and left-handed circular polarizations and we have defined the effective tensor combinations
\begin{equation}
d_{R} = \frac{d_{22}+i d_{31}}{\sqrt{2}},
\qquad
d_{L} = \frac{d_{22}-i d_{31}}{\sqrt{2}}.
\label{eq:d_eff_defs}
\end{equation}

Each term corresponds to a radiated telecom photon with well-defined angular momentum \(m = \pm M \pm \ell N \pm 2\), and polarization \(\ket{R}\) or \(\ket{L}\). These terms determine the angular momentum and polarization of the emitted quantum state.

Because the trigonometric factors in Eq.~\eqref{eq:lnbo_coeffs} decompose entirely into $e^{\pm i2\phi_n}$ harmonics, the anisotropic tensor merely rescales the RHCP/LHCP amplitudes through $d_{R}$ and $d_{L}$ while leaving the selection rules \(M'=\ell N\pm2\) intact. Setting $d_{31}\rightarrow 0$ recovers the isotropic approximation used in the simulations.

The nonlinear polarization $\hat{\mathbf{P}}^{(2)}$ acts as a source term for radiation at the target frequency. Substituting Eq.~(7) into the inhomogeneous wave equation,
\begin{equation}
\Big[
\partial_{\rho\rho} +
\frac{1}{\rho}\partial_{\rho} +
\frac{1}{\rho^{2}} \partial_{\phi\phi} + \partial_{zz}
\Big]\,
\hat{\mathbf{E}}_3
-
\frac{n^{2}}{c^{2}}\,
\partial_{tt}\hat{\mathbf{E}}_3
=
\frac{1}{\varepsilon_0 c^{2}}\,
\partial_{tt}\hat{\mathbf{P}}^{(2)},
\label{supp:eq5}
\end{equation}

We obtain radiation with the same spatiotemporal structure as the nonlinear source. The positive-frequency quantized telecom field reads as

\begin{equation}
\begin{aligned}
\hat{\mathbf{E}}_{3,\mathrm{CW}}^{(+)}(\mathbf r, t)
&=
\sum_{\ell\in\mathbb{Z}}
\Big[
\mathcal{N}_{3,R,\ell}\,J_{Q_{\ell+}}(k_3\rho)\, e^{iQ_{\ell+}\phi}\, e^{\pm i k_{3z} z}\, \hat{a}^{(\mathrm{cw})}_{3,R,\ell}\,\ket{R}\\
&\qquad\quad+
\mathcal{N}_{3,L,\ell}\,J_{Q'_{\ell+}}(k_3\rho)\, e^{iQ'_{\ell+}\phi}\, e^{\pm i k_{3z} z}\, \hat{a}^{(\mathrm{cw})}_{3,L,\ell} \,\ket{L}
\Big] e^{-i\omega_3 t},
\end{aligned}
\label{supp:eq6}
\end{equation}

\begin{equation}
\begin{aligned}
\hat{\mathbf{E}}_{3,\mathrm{CCW}}^{(+)}(\mathbf r, t)
&=
\sum_{\ell\in\mathbb{Z}}
\Big[
\tilde{\mathcal{N}}_{3,R,\ell}\,J_{Q_{\ell-}}(k_3\rho)\, e^{iQ_{\ell-}\phi}\, e^{\pm i k_{3z} z}\, \hat{a}^{(\mathrm{ccw})}_{3,R,\ell}\, \ket{R}\\
&\qquad\quad+
\tilde{\mathcal{N}}_{3,L,\ell}\,J_{Q'_{\ell-}}(k_3\rho)\, e^{-iQ'_{\ell-}\phi}\, e^{\pm i k_{3z} z}\, \hat{a}^{(\mathrm{ccw})}_{3,L,\ell}\, \ket{L}
\Big] e^{-i\omega_3 t},
\end{aligned}
\label{supp:eq7}
\end{equation}

\paragraph*{Interaction Hamiltonian.}
For DFG at $\omega_3=\omega_1-\omega_2$, the $\chi^{(2)}$ interaction Hamiltonian in the rotating wave approximation is

\begin{equation}
\hat{H}_{\mathrm{int}}
= -\frac{1}{2}\int d^3\mathbf{r}\;
\Big[
\hat{\mathbf{E}}_3^{(-)}(\mathbf{r},t)\cdot \hat{\mathbf{P}}^{(2)}(\mathbf{r},t)
+\mathrm{H.c.}
\Big],
\label{supp:eq8}
\end{equation}
where $\hat{\mathbf{E}}_3^{(-)}$ is the negative frequency (creation) part of the telecom field and $\hat{\mathbf{P}}^{(2)}$ is the nonlinear polarization at $\omega_3$ from Eq.~(8). We now substitute the quantized fields into Hamiltonian:

\begin{equation}
\begin{aligned}
\hat{H}_{\mathrm{int}}
&= \hbar\sum_{\ell\in\mathbb{Z}}
\Big(
 g^{(+)}_{R,\ell;\mathrm{CW}}\,\hat{a}^{\dagger}_{3,R,\ell}\,\hat{a}_{1+}
+ g^{(+)}_{L,\ell;\mathrm{CW}}\,\hat{a}^{\dagger}_{3,L,\ell}\,\hat{a}_{1+}\\
&\qquad\quad+
 g^{(+)}_{R,\ell;\mathrm{CCW}}\,\big(\hat{a}^{(\mathrm{ccw})}_{3,R,\ell}\big)^{\dagger}\,\hat{a}_{1+}
+ g^{(+)}_{L,\ell;\mathrm{CCW}}\,\big(\hat{a}^{(\mathrm{ccw})}_{3,L,\ell}\big)^{\dagger}\,\hat{a}_{1+}
\Big)\\
&\quad+\hbar\sum_{\ell\in\mathbb{Z}}
\Big(
 g^{(-)}_{R,\ell;\mathrm{CW}}\,\hat{a}^{\dagger}_{3,R,\ell}\,\hat{a}_{1-}
+ g^{(-)}_{L,\ell;\mathrm{CW}}\,\hat{a}^{\dagger}_{3,L,\ell}\,\hat{a}_{1-}\\
&\qquad\quad+
 g^{(-)}_{R,\ell;\mathrm{CCW}}\,\big(\hat{a}^{(\mathrm{ccw})}_{3,R,\ell}\big)^{\dagger}\,\hat{a}_{1-}
+ g^{(-)}_{L,\ell;\mathrm{CCW}}\,\big(\hat{a}^{(\mathrm{ccw})}_{3,L,\ell}\big)^{\dagger}\,\hat{a}_{1-}
\Big)
+ \text{H.c.}
\end{aligned}
\label{supp:eq9}
\end{equation}

Each coupling $g^{(\pm)}_{s,\ell;\,\mathrm{basis}}$ ($s\!\in\!\{R,L\}$, basis $\in\{\mathrm{CW},\mathrm{CCW}\}$) factors into radial, axial, and angular overlaps:

\begin{equation}
\begin{aligned}
g^{(\pm)}_{s,\ell;\,\mathrm{basis}}
&= g_{0,s}\,
\mathcal{N}_{1}\,
\mathcal{N}^{(\mathrm{basis})}_{3,s,\ell}\,
I_\rho^{(s,\ell;\pm)}\,
I_z^{(\pm)}\,
I_\phi^{(s,\ell;\pm)}.
\end{aligned}
\label{supp:eq10}
\end{equation}
where the channel-dependent prefactors read $g_{0,R}=C_{0}\,d_{R}\alpha^{*}$ and $g_{0,L}=C_{0}\,d_{L}\alpha^{*}$, with $C_{0} \equiv \varepsilon_0 N/(8\pi\hbar)$, and $\mathcal{N}_1$, $\mathcal{N}^{(\mathrm{basis})}_{3,s,\ell}$ are mode normalizations.

\begin{equation}
I_\rho^{(s,\ell;\pm)}
\;\equiv\;
\int_{0}^{\infty} \rho\, d\rho\;
J_M(k_1\rho)\, J_P(k_2\rho)\, J_{m_{s,\ell}^{(\pm)}}(k_3\rho).
\label{eq:radial_overlap}
\end{equation}

\begin{equation}
I_z^{(\pm)}
\;\equiv\;
\int_{-L_z/2}^{L_z/2} dz\; e^{\,i\Delta k_z^{(\pm)} z}
\;=\;
L_z\,\mathrm{sinc}\!\Big(\frac{\Delta k_z^{(\pm)}L_z}{2}\Big)\,
e^{\,i\Delta k_z^{(\pm)}L_z/2}.
\label{eq:axial_overlap}
\end{equation}

\begin{equation}
I_\phi^{(s,\ell;\pm)}
\;\equiv\;
\int_{\phi_0}^{\phi_0+\Phi_0} d\phi\; e^{\,i\Delta m_{s,\ell}^{(\pm)} \phi}
\;=\;
\Phi_0\,\mathrm{sinc}\!\Big(\frac{\Delta m_{s,\ell}^{(\pm)}\Phi_0}{2}\Big)\,
e^{\,i\Delta m_{s,\ell}^{(\pm)}(\phi_0+\Phi_0/2)}.
\label{eq:azimuthal_overlap}
\end{equation}

The axial mismatch is $\Delta k_z^{(\pm)} \equiv \pm k_{1z} + k_{2z} \mp k_{3z}$. We assume the cavity is thin along $z$, so the axial interaction length is short and $\mathrm{sinc}\!\Big(\frac{\Delta m_{s,\ell}^{(\pm)}\Phi_0}{2}\Big) \approx 1$, allowing us to neglect axial phase-mismatch corrections in what follows.

\paragraph*{Selection rules and radiative window.}
For a nonzero interaction, the angular phase matching should be satisfied $\Delta m^{(\pm)}_{s,\ell}=0$, which leads to four selection rules
 
\begin{equation}
\begin{split}
m^{(+)}_{R,\ell}= \ell N + M + 2,\\
m^{(+)}_{L,\ell}= \ell N + M - 2;\\
m^{(-)}_{R,\ell}= \ell N - M + 2,\\ m^{(-)}_{L,\ell}= \ell N - M - 2,
\end{split}
\end{equation}

\paragraph*{Radiative condition.}
The generated mode must be radiative to couple energy out of the cavity. This requires that the wavevector component in the azimuthal direction be smaller than the corresponding free-space wavevector. For RHCP, the radiation condition is: $\big|\dfrac{m}{R}\big| < \dfrac{\omega_3}{c} = \dfrac{2\pi}{\lambda_3}$. Using the relations $M = \dfrac{2\pi R n_{\mathrm{eff}}}{\lambda_1}$, $N = \dfrac{2\pi R}{\Lambda}$, and above selection rules:

For CW, RHCP ($m^{(+)}_{R,\ell}=\ell N+M+2$):
\begin{equation}
-\frac{2\pi R}{\lambda_3} \;<\; \ell N + M + 2 \;<\; \frac{2\pi R}{\lambda_3}
\;\;\Longleftrightarrow\;\;
-\Lambda\!\left(\frac{1}{\lambda_3}+\frac{n_{\mathrm{eff}}}{\lambda_1}+\frac{1}{\pi R}\right)
\;<\;
\ell
\;<\;
\Lambda\!\left(\frac{1}{\lambda_3}-\frac{n_{\mathrm{eff}}}{\lambda_1}-\frac{1}{\pi R}\right).
\label{eq:ell_window_CW_R}
\end{equation}

For CW, LHCP ($m^{(+)}_{L,\ell}=\ell N+M-2$):
\begin{equation}
-\frac{2\pi R}{\lambda_3} \;<\; \ell N + M - 2 \;<\; \frac{2\pi R}{\lambda_3}
\;\;\Longleftrightarrow\;\;
\Lambda\!\left(-\frac{n_{\mathrm{eff}}}{\lambda_1}-\frac{1}{\lambda_3}+\frac{1}{\pi R}\right)
\;<\;
\ell
\;<\;
\Lambda\!\left(-\frac{n_{\mathrm{eff}}}{\lambda_1}+\frac{1}{\lambda_3}+\frac{1}{\pi R}\right).
\label{eq:ell_window_CW_L}
\end{equation}

For CCW, RHCP ($m^{(-)}_{R,\ell}=\ell N - M + 2$):
\begin{equation}
-\frac{2\pi R}{\lambda_3} \;<\; \ell N - M + 2 \;<\; \frac{2\pi R}{\lambda_3}
\;\;\Longleftrightarrow\;\;
\Lambda\!\left(-\frac{1}{\lambda_3}+\frac{n_{\mathrm{eff}}}{\lambda_1}-\frac{1}{\pi R}\right)
\;<\;
\ell
\;<\;
\Lambda\!\left(\frac{1}{\lambda_3}+\frac{n_{\mathrm{eff}}}{\lambda_1}-\frac{1}{\pi R}\right).
\label{eq:ell_window_CCW_R}
\end{equation}

For CCW, LHCP ($m^{(-)}_{L,\ell}=\ell N - M - 2$):
\begin{equation}
-\frac{2\pi R}{\lambda_3} \;<\; \ell N - M - 2 \;<\; \frac{2\pi R}{\lambda_3}
\;\;\Longleftrightarrow\;\;
\Lambda\!\left(-\frac{1}{\lambda_3}+\frac{n_{\mathrm{eff}}}{\lambda_1}+\frac{1}{\pi R}\right)
\;<\;
\ell
\;<\;
\Lambda\!\left(\frac{1}{\lambda_3}+\frac{n_{\mathrm{eff}}}{\lambda_1}+\frac{1}{\pi R}\right).
\label{eq:ell_window_CCW_L}
\end{equation}

With the CW case and parameters
$(M,N,\lambda_3)=(21,23,1350~\mathrm{nm})$, one finds $\ell=-1$ as the only allowed order, hence the OAM of radiative channels are: 
\begin{equation}
\hbar m^{(+)}_{R}=\hbar\,(M-N+2),\qquad
\hbar m^{(+)}_{L}=\hbar\,(M-N-2).
\label{eq:OAM_CW_example}
\end{equation}
Analogously, for CCW channels the allowed order is $\ell=1$ from

\begin{equation}
\hbar m^{(-)}_{R}=\hbar\,(-M+N+2),\qquad
\hbar m^{(-)}_{L}=\hbar\,(-M+N-2).
\label{eq:OAM_CCW_example}
\end{equation}

\paragraph*{Dominant diffraction orders.}
The Hamiltonian for these orders is

\begin{equation}
\hat{H}_{\mathrm{int}}^{(\mathrm{CW})}
=
\hbar\Big[
g^{(+)}_{R;\mathrm{CW}}\,
\big(\hat{a}^{(\mathrm{cw})}_{3,R}\big)^{\dagger}\,\hat{a}_{1+}
+
g^{(+)}_{L;\mathrm{CW}}\,
\big(\hat{a}^{(\mathrm{cw})}_{3,L}\big)^{\dagger}\,\hat{a}_{1+}
\Big]
+\mathrm{H.c.}
\label{eq:Hcw_lm1}
\end{equation}

and,

\begin{equation}
\hat{H}_{\mathrm{int}}^{(\mathrm{CCW})}
=
\hbar\Big[
g^{(-)}_{R;\mathrm{CCW}}\,
\big(\hat{a}^{(\mathrm{ccw})}_{3,R}\big)^{\dagger}\,\hat{a}_{1-}
+
g^{(-)}_{L;\mathrm{CCW}}\,
\big(\hat{a}^{(\mathrm{ccw})}_{3,L}\big)^{\dagger}\,\hat{a}_{1-}
\Big]
+\mathrm{H.c.}
\label{eq:Hccw_lp1}
\end{equation}

\paragraph*{Bright supermodes and conversion efficiency.}
We define for each propagation direction, a \emph{bright} supermode as the coupling-weighted sum over the two (RHCP/LHCP) radiative channels of the dominant diffraction order:
\begin{equation}
\hat{b}_{+}
\;\equiv\;
\frac{ g^{(+)}_{R;\mathrm{CW}}\;\hat{a}^{(\mathrm{cw})}_{3,R}
      + g^{(+)}_{L;\mathrm{CW}}\;\hat{a}^{(\mathrm{cw})}_{3,L} }
     { g_{\mathrm{eff}}^{(+)} },
\qquad
g_{\mathrm{eff}}^{(+)} \;\equiv\;
\sqrt{ \big|g^{(+)}_{R;\mathrm{CW}}\big|^2 + \big|g^{(+)}_{L;\mathrm{CW}}\big|^2 },
\label{eq:bright_cw}
\end{equation}
\begin{equation}
\hat{b}_{-}
\;\equiv\;
\frac{ g^{(-)}_{R;\mathrm{CCW}}\;\hat{a}^{(\mathrm{ccw})}_{3,R}
      + g^{(-)}_{L;\mathrm{CCW}}\;\hat{a}^{(\mathrm{ccw})}_{3,L} }
     { g_{\mathrm{eff}}^{(-)} },
\qquad
g_{\mathrm{eff}}^{(-)} \;\equiv\;
\sqrt{ \big|g^{(-)}_{R;\mathrm{CCW}}\big|^2 + \big|g^{(-)}_{L;\mathrm{CCW}}\big|^2 }.
\label{eq:bright_ccw}
\end{equation}

The CW/CCW WGMs have orthonormal spatial profiles; hence
\(
[\hat a_{1+},\,\hat a_{1-}^{\dagger}] = 0.
\label{eq:comm_WGM}
\)

Moreover, the radiative modes are represented in cylindrical vector harmonics. Therefore, the modes with different azimuthal order $m$ and circular polarization $s\in\{R,L\}$, are orthogonal. Meaning that
\(
[\hat a_{3,(m,s)},\,\hat a_{3,(m',s')}^{\dagger}] \;=\; \delta_{m,m'}\,\delta_{s,s'}.
\label{eq:comm_fs_basis}
\)
The four radiative operators used above are specific members of this basis, therefore,

\begin{equation}
\big[\hat a^{(\mathrm{cw})}_{3,s},\,\hat a^{(\mathrm{cw})\dagger}_{3,s'}\big] = \delta_{s,s'}, 
\qquad
\big[\hat a^{(\mathrm{ccw})}_{3,s},\,\hat a^{(\mathrm{ccw})\dagger}_{3,s'}\big] = \delta_{s,s'},
\label{eq:comm_same_dir}
\end{equation}
\begin{equation}
\big[\hat a^{(\mathrm{cw})}_{3,s},\,\hat a^{(\mathrm{ccw})\dagger}_{3,s'}\big] 
= 0 \quad \text{for } m^{(+)}_{s}\neq m^{(-)}_{s'}.
\label{eq:comm_cross_dir}
\end{equation}
Equations \eqref{eq:comm_same_dir} to \eqref{eq:comm_cross_dir} imply, for the bright supermodes defined in Eqs.~\eqref{eq:bright_cw} to \eqref{eq:bright_ccw},
\begin{equation}
[\hat b_{\pm},\hat b_{\pm}^{\dagger}]=1, \qquad [\hat b_{+},\hat b_{-}^{\dagger}]=0.
\label{eq:comm_bright}
\end{equation}

Using supermodes, the total interaction Hamiltonian reduces to a pair of coupled modes:

\begin{equation}
\hat{H}_{\mathrm{int}}
=
\hbar\,g_{\mathrm{eff}}^{(+)}\Big( \hat{b}_{+}^{\dagger}\hat{a}_{1+} + \hat{a}_{1+}^{\dagger}\hat{b}_{+} \Big)
\;+\;
\hbar\,g_{\mathrm{eff}}^{(-)}\Big( \hat{b}_{-}^{\dagger}\hat{a}_{1-} + \hat{a}_{1-}^{\dagger}\hat{b}_{-} \Big).
\label{eq:H_bright}
\end{equation}


Using the calculated Hamiltonian, the Heisenberg equations are obtained as follows:
\begin{equation}
\dot{\hat{a}}_{1+}=-i g_{\mathrm{eff}}^{(+)}\hat{b}_{+},\quad
\dot{\hat{b}}_{+}=-i g_{\mathrm{eff}}^{(+)}\hat{a}_{1+},
\qquad
\dot{\hat{a}}_{1-}=-i g_{\mathrm{eff}}^{(-)}\hat{b}_{-},\quad
\dot{\hat{b}}_{-}=-i g_{\mathrm{eff}}^{(-)}\hat{a}_{1-},
\end{equation}

Solving these equations gives Rabi oscillations between the color-center-coupled WGM and the corresponding bright output mode. 
The solutions describe sinusoidal exchange between the modes:
\begin{equation}
\hat{a}_{1,\pm}(t) = \hat{a}_{1,\pm}(0)\cos(g^{\pm}_{\text{eff}}t) - i\hat{b}_{\pm}(0)\sin(g^{\pm}_{\text{eff}}t), 
\qquad
\hat{b}_{\pm}(t) = \hat{b}_{\pm}(0)\cos(g^{\pm}_{\text{eff}}t) - i\hat{a}_{\pm}(0)\sin(g^{\pm}_{\text{eff}}t).
\label{eq:heisenberg_solution}
\end{equation}

If the interaction starts with one photon in the WGM and vacuum in the bright mode,
\(
\langle \hat{a}^\dagger \hat{a} \rangle(0) = 1, 
\text{and}\, 
\langle \hat{b}^\dagger \hat{b} \rangle(0) = 0,
\)
then the conversion probability to the bright mode is

\begin{equation}
\eta^{(\pm)} = \sin^2\!\big(g_{\mathrm{eff}}^{(\pm)} T\big),
\label{eq:eta_rabi}
\end{equation}
where $T$ is the interaction time (or equivalently, the azimuthal propagation length $g_{\mathrm{eff}}T \to \gamma_{\phi} L_{\phi}$). Thus, where
\(
\gamma_{\phi} L_{\phi} = g_{\mathrm{eff}} T,
\text{and}\,
\gamma_\phi=\frac{g_{\mathrm{eff}}}{v_\phi},
\).

Within each direction, the mean photon numbers in the two physical polarization channels are represented by:
\begin{equation}
\langle \hat{n}^{(\mathrm{cw})}_{3,R} \rangle
=
\frac{ \big|g^{(+)}_{R;\mathrm{CW}}\big|^{2} }{ \big(g_{\mathrm{eff}}^{(+)}\big)^{2} }\;\eta^{(+)}\;\langle \hat{n}_{1+}(0)\rangle,
\quad
\langle \hat{n}^{(\mathrm{cw})}_{3,L} \rangle
=
\frac{ \big|g^{(+)}_{L;\mathrm{CW}}\big|^{2} }{ \big(g_{\mathrm{eff}}^{(+)}\big)^{2} }\;\eta^{(+)}\;\langle \hat{n}_{1+}(0)\rangle,
\label{eq:split_cw}
\end{equation}
\begin{equation}
\langle \hat{n}^{(\mathrm{ccw})}_{3,R} \rangle
=
\frac{ \big|g^{(-)}_{R;\mathrm{CCW}}\big|^{2} }{ \big(g_{\mathrm{eff}}^{(-)}\big)^{2} }\;\eta^{(-)}\;\langle \hat{n}_{1-}(0)\rangle,
\quad
\langle \hat{n}^{(\mathrm{ccw})}_{3,L} \rangle
=
\frac{ \big|g^{(-)}_{L;\mathrm{CCW}}\big|^{2} }{ \big(g_{\mathrm{eff}}^{(-)}\big)^{2} }\;\eta^{(-)}\;\langle \hat{n}_{1-}(0)\rangle.
\label{eq:split_ccw}
\end{equation}

\subsection*{B. Write-in process}

The write-in operation performs the reverse mapping: a telecom photon injected through the fiber mode is upconverted into the WGM at $\omega_1$ using the same programmable pump, now in the sum-frequency generation (SFG) configuration with $\omega_1 = \omega_2 + \omega_3$ ($\omega_3 > \omega_2$). As above, we treat the pump as a classical drive and the signal modes as quantized, and we mirror the three-step structure of the read-out derivation.

\paragraph*{Field definitions.}
To maximise spatial overlap, both the pump and the fiber excitations carry the $N$-fold angular modulation. Each field illuminates the resonator from the $-z$ direction, with the pump taken as strong and undepleted while the fiber mode remains quantized. Under these assumptions the structured pump and quantized fiber fields are

\begin{equation}
\begin{split}
\mathbf E_2^{(+)}(\mathbf r,t)
&= \alpha\, J_N^2(k_2 \rho)
\sum_{n=0}^{N-1}
\delta(\phi-\phi_n)
 e^{-i k_{2z} z} e^{-i\omega_2 t}
\hat{\boldsymbol{\rho}}_{n},\\[3pt]
\hat{\mathbf E}_3^{(+)}(\mathbf r,t)
&= \mathcal N_3\, \hat a_3\, J_N^2(k_3 \rho)
\sum_{n=0}^{N-1}
\delta(\phi-\phi_n)
 e^{-i k_{3z} z} e^{-i\omega_3 t}
\hat{\boldsymbol{\phi}}_{n},
\end{split}
\label{eq:SFG_fields}
\end{equation}
where $\alpha$ is the classical pump amplitude, $\mathcal N_3$ is the field normalization for the quantized fiber mode, and $\hat a_3$ is its annihilation operator. The spatial function $J_N(\rho,z)$ represents the radial and axial mode overlap at the $N$ discrete azimuthal positions $\phi_n = 2\pi n / N$.

\paragraph*{Nonlinear polarization and cavity response.}
The second-order nonlinear polarization at $\omega_1 = \omega_2 + \omega_3$ is operator-valued and given by
\begin{equation}
\hat{\mathbf P}^{(2)(+)}(\mathbf r,t;\omega_1)
= \varepsilon_0\, \overline{\overline{\chi}}^{(2)} :
\big[
\mathbf E_{2}^{(+)}(\mathbf r,t)
\hat{\mathbf E}_{3}^{(+)}(\mathbf r,t)
\big].
\label{eq:SFG_P2_def}
\end{equation}

For LiNbO$_3$, only the $\chi_{22}$ component contributes efficiently for the chosen polarization geometry (pump: $\hat{\boldsymbol{\phi}}$; fiber: $\hat{\boldsymbol{\rho}}$; TE-like WGM: transverse $\hat{\boldsymbol{\rho}}$). Evaluating the tensor projection gives a circularly polarized Fourier series:
\begin{equation}
\begin{aligned}
\hat{\mathbf P}^{(2)(+)}(\mathbf r,t;\omega_1)
&=
-\frac{N}{4\pi}\, \varepsilon_0\,\chi_{22}\,
\mathcal N_3\,\alpha
 e^{-i (k_{2z}+k_{3z}) z} e^{-i\omega_1 t}
\hat a_3
\sum_{\ell\in\mathbb{Z}}
\Big[
 e^{i(\ell N + 2)\phi}\,\ket{R}
+ e^{i(\ell N - 2)\phi}\,\ket{L}
\Big]\\
&\qquad\times
 J_N(k_2 \rho)\,
 J_N(k_3 \rho),
\end{aligned}
\label{eq:SFG_P2_series}
\end{equation}

The polarization generates a new field at $\omega_1$ as
\begin{equation}
\begin{aligned}
\hat{\mathbf E}_{1}^{(+)}(\mathbf r,t)
&=
\mathcal N_{1} e^{\pm i k_{1z}z}
 e^{-i\omega_{1} t}
\sum_{\ell\in\mathbb{Z}}
\Big[
J_{M^{(\text{cw})}_{R,\ell}}(k_{1}\rho)\,
\hat a^{(\text{cw})}_{1,R,\ell}\,
 e^{+iM^{(\text{cw})}_{R,\ell}\phi}\ket{R}\\
&\qquad+
J_{M^{(\text{cw})}_{L,\ell}}(k_{1}\rho)\,
\hat a^{(\text{cw})}_{1,L,\ell}\,
 e^{+iM^{(\text{cw})}_{L,\ell}\phi}\ket{L}\\
&\qquad+
J_{M^{(\text{ccw})}_{R,\ell}}(k_{1}\rho)\,
\hat a^{(\text{ccw})}_{1,R,\ell}\,
 e^{-iM^{(\text{ccw})}_{R,\ell}\phi}\ket{R}\\
&\qquad+
J_{M^{(\text{ccw})}_{L,\ell}}(k_{1}\rho)\,
\hat a^{(\text{ccw})}_{1,L,\ell}\,
 e^{-iM^{(\text{ccw})}_{L,\ell}\phi}\ket{L}
\Big],
\end{aligned}
\label{eq:SFG_E1_full}
\end{equation}
where $\hat a_{1\pm}$ are the annihilation operators for the clockwise (CW) and counterclockwise (CCW) WGMs, respectively, both sharing the azimuthal index $M$ and transverse polarization $\hat{\boldsymbol{\rho}}$.

\paragraph*{Effective Hamiltonian and bright mode.}
The SFG interaction Hamiltonian at $\omega_1=\omega_2+\omega_3$ is
\begin{equation}
\hat H_{\text{SFG}}=
\hbar\sum_{\ell\in\mathbb Z}\sum_{s\in\{R,L\}}
\Big[
G_{\text{cw},s,\ell}\,\big(\hat a^{(\mathrm{cw})}_{1,s,\ell}\big)^{\dagger}\hat a_{3}
+
G_{\text{ccw},s,\ell}\,\big(\hat a^{(\mathrm{ccw})}_{1,s,\ell}\big)^{\dagger}\hat a_{3}
\Big]
+\mathrm{H.c.},
\label{eq:SFG_H_sum_full_revised}
\end{equation}
where $s\in\{R,L\}$ labels circular polarization.

Each coupling coefficient factors into radial, axial, and angular parts:
\begin{equation}
\begin{split}
G_{\text{cw},s,\ell}
&=
G_0\;\mathcal N_1\mathcal N_3\;
I_{\rho}^{(\text{cw},s,\ell)}\;I_{z}\;I_{\phi}^{(\text{cw},s)}(\ell),\\
G_{\text{ccw},s,\ell}
&=
G_0\;\mathcal N_1\mathcal N_3\;
I_{\rho}^{(\text{ccw},s,\ell)}\;I_{z}\;I_{\phi}^{(\text{ccw},s)}(\ell),
\label{eq:SFG_G_def_full_revised}
\end{split}
\end{equation}
where $G_0\equiv \frac{\varepsilon_0\chi_{22}\,\alpha}{2\hbar}\,\frac{N}{4\pi}$.
The radial overlap integrals are
\begin{equation}
\begin{split}
I_{\rho}^{(\mathrm{cw},s,\ell)}
&=\int_{0}^{\infty}\rho\,d\rho\;
J_{M^{(\mathrm{cw})}_{s,\ell}}(k_1\rho)\,J_N(k_2\rho)\,J_N(k_3\rho),\\
I_{\rho}^{(\mathrm{ccw},s,\ell)}
&=\int_{0}^{\infty}\rho\,d\rho\;
J_{M^{(\mathrm{ccw})}_{s,\ell}}(k_1\rho)\,J_N(k_2\rho)\,J_N(k_3\rho),
\label{eq:SFG_Irho_dir_spin}
\end{split}
\end{equation}
with axial factor
\begin{equation}
I_{z}=\int_{-L_z/2}^{L_z/2}\!dz\;e^{\,i\Delta k_z z}=L_z\,\mathrm{sinc}\!\Big(\tfrac{\Delta k_z L_z}{2}\Big)\,e^{\,i\Delta k_z L_z/2},
\label{eq:SFG_Iz_revised}
\end{equation}
where $\Delta k_z= \pm k_{1z}-(k_{2z}+k_{3z})$ is the axial phase mismatch. 

The angular overlaps are
\begin{equation}
\begin{split}
I_{\phi}^{(\mathrm{cw},s)}(\ell)&=\int_{0}^{2\pi}\!d\phi\;e^{\,i[(\ell N+\sigma_s 2)-M]\phi}=2\pi\,\delta_{M,\;\ell N+\sigma_s 2},\\
I_{\phi}^{(\mathrm{ccw},s)}(\ell)&=\int_{0}^{2\pi}\!d\phi\;e^{\,i[(\ell N+\sigma_s 2)+M]\phi}=2\pi\,\delta_{M,\;-\ell N-\sigma_s 2},
\label{eq:SFG_Iphi_full_revised}
\end{split}
\end{equation}
with $\sigma_s=1$ for RHCP and $\sigma_s=-1$ for LHCP.

Non-zero coupling requires the angular phase-matching conditions
\begin{equation}
\begin{split}
M^{(\mathrm{cw})}_{s,\ell}=\ell N+\sigma_s 2\ge 0,\\
M^{(\mathrm{ccw})}_{s,\ell}=-(\ell N+\sigma_s 2)\ge 0,
\end{split}
\end{equation}
so that $M_{s,\ell}=\big|\ell N+\sigma_s 2\big|$.

Guided WGMs at $\omega_1$ satisfy $k_0 n_{\rm out}<M_{s,\ell}/R<k_0 n_{\rm core}$ with $k_0=2\pi/\lambda_1$, yielding
\begin{equation}
\frac{k_0 n_{\rm out}R-\sigma_s 2}{N}<\ell<\frac{k_0 n_{\rm core}R-\sigma_s 2}{N},
\label{eq:ell_window_CW}
\end{equation}
for the CW branch and
\begin{equation}
\frac{-k_0 n_{\rm core}R-\sigma_s 2}{N}<\ell<\frac{-k_0 n_{\rm out}R-\sigma_s 2}{N},
\label{eq:ell_window_CCW}
\end{equation}
for the CCW branch. For $N=23$, $R=1.66\,\mu\mathrm{m}$, $n_{\text{core}}=2.4$, $n_{\text{out}}=1$, and $\lambda_1 = 737\,\mathrm{nm}$, the dominant solutions are $\ell=1$ (CW) and $\ell=-1$ (CCW), giving
\begin{equation}
\begin{split}
M_{R}^{(\mathrm{cw})}=M_{R}^{(\mathrm{ccw})} = N + 2,\\
M_{L}^{(\mathrm{cw})}=M_{L}^{(\mathrm{ccw})} = N - 2.
\end{split}
\end{equation}

The interaction Hamiltonian becomes
\begin{equation}
\begin{aligned}
\hat H_{\text{SFG}}
&= \hbar\Big[
G_{\mathrm{cw},R}\,\big(\hat a_{1,\mathrm{cw},R}\big)^{\dagger}\hat a_{3}
+ G_{\mathrm{cw},L}\,\big(\hat a_{1,\mathrm{cw},L}\big)^{\dagger}\hat a_{3}\\
&\qquad\quad+
G_{\mathrm{ccw},R}\,\big(\hat a_{1,\mathrm{ccw},R}\big)^{\dagger}\hat a_{3}
+ G_{\mathrm{ccw},L}\,\big(\hat a_{1,\mathrm{ccw},L}\big)^{\dagger}\hat a_{3}
\Big] + \mathrm{H.c.}
\end{aligned}
\label{eq:H_SFG_single_input}
\end{equation}

Defining the cavity bright supermode
\begin{equation}
\hat b=\frac{1}{g_{\mathrm{eff}}}\sum_{j=1}^{4}G_j^{*}\,\hat a_{1,j},\qquad g_{\mathrm{eff}}=\Big(\sum_{j=1}^{4}|G_j|^{2}\Big)^{1/2},
\label{eq:bright_single_input}
\end{equation}
with $[\hat b,\hat b^{\dagger}]=1$ casts the Hamiltonian as
\begin{equation}
\hat H_{\text{SFG}}=\hbar\,g_{\mathrm{eff}}\big(\hat b^{\dagger}\hat a_{3}+\hat a_{3}^{\dagger}\hat b\big).
\label{eq:H_bright_single}
\end{equation}

The Heisenberg equations
\begin{equation}
\dot{\hat b}=-i\,g_{\mathrm{eff}}\,\hat a_{3},\qquad\dot{\hat a}_{3}=-i\,g_{\mathrm{eff}}\,\hat b
\label{eq:EOM_b_a3_single}
\end{equation}
lead to
\begin{equation}
\begin{split}
\hat a_{3}(t)=\hat a_{3}(0)\cos(g_{\mathrm{eff}}t)-i\,\hat b(0)\sin(g_{\mathrm{eff}}t),\\
\hat b(t)=\hat b(0)\cos(g_{\mathrm{eff}}t)-i\,\hat a_{3}(0)\sin(g_{\mathrm{eff}}t).
\end{split}
\end{equation}
Assuming an initially empty cavity, $\hat b(0)|\psi_0\rangle=0$, and $n_{\rm in}=\langle \hat a_{3}^{\dagger}(0)\hat a_{3}(0)\rangle$, the internal SFG efficiency is
\begin{equation}
\eta_{\mathrm{SFG}}^{(\mathrm{int})}\equiv\frac{\langle \hat b^{\dagger}(T)\hat b(T)\rangle}{n_{\rm in}}=\sin^{2}\!\big(g_{\mathrm{eff}}T\big).
\label{eq:eta_single_input}
\end{equation}
Population in each cavity channel follows
\begin{equation}
\langle \hat a_{1,j}^{\dagger}(T)\hat a_{1,j}(T)\rangle=\Big|\frac{G_j}{g_{\mathrm{eff}}}\Big|^{2}\,\eta_{\mathrm{SFG}}^{(\mathrm{int})}\,n_{\rm in},\qquad j=1,\dots,4,
\label{eq:partition_single_input}
\end{equation}
and for azimuthally distributed interactions one replaces $g_{\mathrm{eff}}T$ with $\gamma_{\phi}L_{\phi}$. In that case

\begin{equation}
\eta_{\mathrm{SFG}}^{(\mathrm{int})}=\sin^{2}(\gamma_{\phi}L_{\phi}).
\end{equation}

\section*{II. Far-field emission}
\phantomsection\label{supp:secII}

We now connect the nonlinear polarization of Sec.~I to the radiated field collected in the far zone. Each radiative channel $j\in\{R{+},L{+},R{-},L{-}\}$ carries a definite circular polarization $\sigma_j\in\{R,L\}$ and azimuthal order $Q_j\in\{Q_{+},Q^{\prime}_{+},Q_{-},Q^{\prime}_{-}\}$. The analysis proceeds in three steps: we first cast the $N$ pump sites as a phased dipole array, then evaluate the far-field Green-function integral, and finally assemble the intensity including the relative populations and coherences dictated by the bright-mode Hamiltonian.

\subsection*{A. Array factor and Green functions}

\paragraph*{Dipole array factor.}
The resonator is thin along $z$ with interaction length $L_z$, so each azimuthal site is modeled as a virtual dipole emitting at $\omega_3$. The spatio-temporal form of the emission from channel $j$ at position $\phi_n$ is

\begin{equation}
\hat{\mathbf E}^{(+)}_{j,n}(t)
=
\mathcal N_j\,\hat a_{3,j}\,e^{iQ_j\phi_n}\,e^{-i\omega_3 t}\,|\sigma_j\rangle.
\label{eq:45}
\end{equation}
Here $\mathcal N_j$ is the total normalization.

The positive-frequency far field for channel \(j\) is obtained using Dyadic Green's function

\begin{equation}
\hat{\mathbf E}^{(+)}_{\text{ff},j}(\mathbf r,t)
=
\frac{k_3^2\, \chi_{\text{eff}} \, e^{ik_3 r}}{4\pi r}\;
\big(\overline{\overline I}-\hat{\mathbf r}\hat{\mathbf r}\big)\,\sum_{n=1}^{N}
\int d^3 r'\;\hat{\mathbf E}^{(+)}_{j,n}(\mathbf r',t)\,e^{-ik_3\hat{\mathbf r}\!\cdot\!\mathbf r'} .
\label{eq:green_start}
\end{equation}
where \(\chi_{\text{eff}}\) is the linear susceptibility. We find farfield for radiative channel with OAM of \(\hbar m^{(+)}_{R} \). Farfield of other channels is calculated in a similar way.  

By simplification, the azimuthal array factor is expressed as

\begin{equation}
S_{m^{(+)}_{R} }(\theta, \phi) =
\sum_{n=1}^{N}
e^{im^{(+)}_{R} \phi_n}\, e^{-ik_3 R \sin \theta \cos(\phi_n-\phi)}.
\label{eq:array_factor}
\end{equation}
In the dense sampling limit ($N\to\infty$), using the Jacobi Anger expansion and uniform sampling gives
\begin{equation}
S_{m^{(+)}_{R} }(\theta, \phi)
\underset{N\to\infty}{\longrightarrow}
N\, (-i)^{m^{(+)}_{R}}\, e^{im^{(+)}_{R} \phi}\,
J_{m^{(+)}_{R} }\!\left(k_3 R \sin \theta\right).
\label{eq:array_factor_dense}
\end{equation}

\paragraph*{Polarization content.}
We next evaluate the polarization of the radiated field. For
the RHCP polarization $\ket{R} = \dfrac{\hat x+i\hat y}{\sqrt{2}}$:
\begin{equation}
\big[ \overline{\overline{I}} - \hat r \hat r \big]\ket{R}
= \ket{R}
- \frac{\sin \theta\, e^{i\phi}}{\sqrt{2}}
\Big(
\sin \theta \cos\phi\, \hat x + \sin \theta \sin\phi\, \hat y + \cos \theta\, \hat z
\Big).
\label{eq:projection_R}
\end{equation}
\paragraph*{Far-field envelope.}
In the paraxial cone ($\theta \ll 1$), the projection yields the RHCP Jones vector $\ket{R}$ up to corrections $\mathcal{O}(\theta^{2})$. The resulting field amplitude is
\begin{equation}
E_{\mathrm{m^{(+)}_{R}}} = N\, g(\theta, r)\, (-i)^{m^{(+)}_{R}}\, e^{im^{(+)}_{R}\phi}\,
J_{m^{(+)}_{R}}\!\left(k_3 R \sin \theta\right)\, \ket{R}
\label{eq:farfield_envelope}
\end{equation}
With paraxial approximation, $r \simeq z + \dfrac{R^{2}+\rho^{2}}{2z}$
and $\sin \theta \simeq \tan \theta = \rho/z$, the free space propagator is defined as
\begin{equation}
g(\theta, r) =
\frac{P_0\, \omega_3^{2}}{4\pi\varepsilon_0 c^{2}}\,
\frac{e^{ik_3\left(z+\frac{R^{2}+\rho^{2}}{2z}\right)}}{z}\, L_{z},
\label{eq:free_space_propagator}
\end{equation}

Summing over the four channels gives the total positive-frequency far field
$\hat{\mathbf E}^{(+)}_{\rm ff}=\sum_{j}\hat{\mathbf E}^{(+)}_{\rm ff,\,j}$.

\paragraph*{Operator and normalization dictionary.}
For quick reference, Table~\ref{tab:operators} lists the annihilation operators that appear in the main text and their physical meaning. All radiative operators obey the usual bosonic algebra, e.g.\ $[\hat a_{3,R}^{(\mathrm{cw})},\hat a_{3,R}^{(\mathrm{cw})\dagger}]=1$, with vanishing commutators between distinct channels. The normalization factor introduced in Eq.~\eqref{eq:qI_explicit} is
\begin{equation}
\mathcal N = \frac{k_3^2 \chi_{\mathrm{eff}}}{4\pi r}\,e^{ik_3 r},
\qquad
g(\theta,r) = \frac{P_0\, \omega_3^{2}}{4\pi\varepsilon_0 c^{2}}\,
\frac{e^{ik_3\left(z+\frac{R^{2}+\rho^{2}}{2z}\right)}}{z}\, L_{z},
\label{eq:norm_summary}
\end{equation}
so that $|\mathcal N g|^2$ reproduces the free-space Green-function envelope used in Eq.~\eqref{eq:free_space_propagator}. The effective nonlinearity $\chi_{\mathrm{eff}}$ includes the tensor contraction with the pump and signal polarizations, and $P_0$ carries the pump-power normalization.

\begin{figure}[H]
  \centering
  \includegraphics{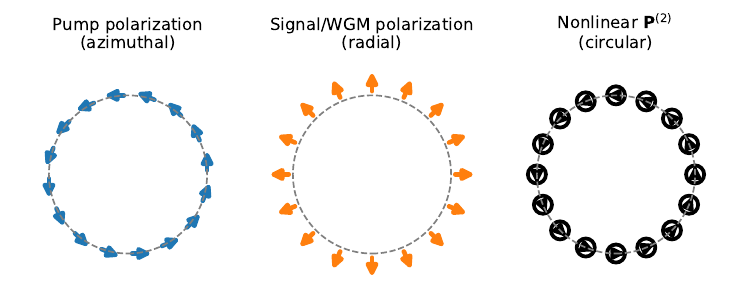}\\[6pt]
  \includegraphics{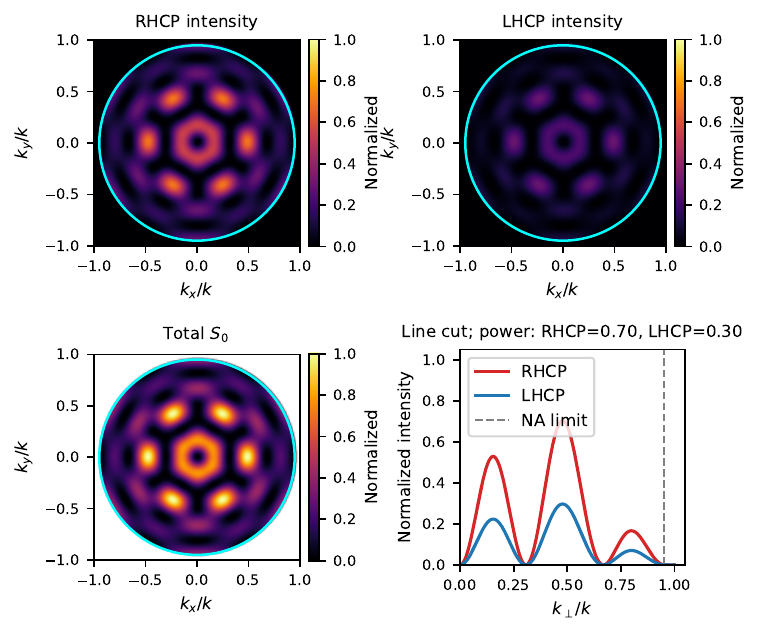}
  \caption{Polarization pathway (theory/simulation). Top: focal-plane polarization example in which the pump is azimuthal, the WGM/signal is radial, and their \(\chi^{(2)}\) product produces a circular nonlinear polarization on the ring. Bottom: far-field RHCP/LHCP maps (analytic model with \(M=20\), \(N=17\), \(\ell=1\)) and line cut showing the R/L power split within the objective NA (cyan circle).}
  \label{fig:polarization-story}
\end{figure}
\FloatBarrier

\begin{table}[t]
\centering
\begin{tabular}{lll}
\hline\hline
Symbol & Physical channel & Notes \\
\hline
$\hat a_{3,R}^{(\mathrm{cw})}$ & RHCP, clockwise radiation & $[\hat a,\hat a^\dagger]=1$ \\
$\hat a_{3,L}^{(\mathrm{cw})}$ & LHCP, clockwise radiation & orthogonal to RHCP \\
$\hat a_{3,R}^{(\mathrm{ccw})}$ & RHCP, counterclockwise radiation &  \\
$\hat a_{3,L}^{(\mathrm{ccw})}$ & LHCP, counterclockwise radiation &  \\
$\hat b_{+}$ & CW bright supermode & Eq.~\eqref{eq:bright_single_input} \\
$\hat b_{-}$ & CCW bright supermode & Eq.~\eqref{eq:bright_single_input} \\
\hline\hline
\end{tabular}
\caption{Operator summary for the radiative channels and their bright-mode combinations.}
\label{tab:operators}
\end{table}

\subsection*{B. Intensity and coherence matrices}

\paragraph*{Intensity and coherence.}
The farfield intensity is
\(
I(\theta,\phi)
=\big\langle \hat{\mathbf E}^{(-)}_{\text{ff}}\!\cdot\!\hat{\mathbf E}^{(+)}_{\text{ff}}\big\rangle\)

Because $\langle R|L\rangle=0$, RHCP and LHCP do not interfere. Group the RHCP pair
$\{i=R{+},R{-}\}$ into
\begin{equation}
\mathbf v_R =
\begin{bmatrix}
\tilde S_{Q_{+}}\\[2pt]
\tilde S_{Q_{-}}
\end{bmatrix},
\qquad
\Gamma_R =
\begin{bmatrix}
\langle \hat a_{3,R{+}}^{\dagger}\hat a_{3,R{+}}\rangle &
\langle \hat a_{3,R{+}}^{\dagger}\hat a_{3,R{-}}\rangle\\[4pt]
\langle \hat a_{3,R{-}}^{\dagger}\hat a_{3,R{+}}\rangle &
\langle \hat a_{3,R{-}}^{\dagger}\hat a_{3,R{-}}\rangle
\end{bmatrix},
\end{equation}
where $\tilde S_i = \mathcal{N}_i\, g\, S_{Q_i}$.
Similarly,
\begin{equation}
\mathbf v_L =
\begin{bmatrix}
\tilde S_{Q'_{+}}\\[2pt]
\tilde S_{Q'_{-}}
\end{bmatrix},
\qquad
\Gamma_L =
\begin{bmatrix}
\langle \hat a_{3,L{+}}^{\dagger}\hat a_{3,L{+}}\rangle &
\langle \hat a_{3,L{+}}^{\dagger}\hat a_{3,L{-}}\rangle\\[4pt]
\langle \hat a_{3,L{-}}^{\dagger}\hat a_{3,L{+}}\rangle &
\langle \hat a_{3,L{-}}^{\dagger}\hat a_{3,L{-}}\rangle
\end{bmatrix}.
\end{equation}
Then
\begin{equation}
I(\theta,\phi)
=\mathbf v_R^{\dagger}\Gamma_R \mathbf v_R + \mathbf v_L^{\dagger}\Gamma_L \mathbf v_L.
\label{eq:qI_quad}
\end{equation}
Writing $n_{R\pm}=\langle \hat a_{3,R\pm}^{\dagger}\hat a_{3,R\pm}\rangle$,
$n_{L\pm}=\langle \hat a_{3,L\pm}^{\dagger}\hat a_{3,L\pm}\rangle$ and coherences
$\Gamma_{R}=\langle \hat a_{3,R{+}}^{\dagger}\hat a_{3,R{-}}\rangle=|\Gamma_{R}|e^{i\psi_R}$,
$\Gamma_{L}=|\Gamma_{L}|e^{i\psi_L}$, and using
$S_{Q}(\theta,\phi)=(-i)^{Q}e^{iQ\phi}J_{Q}(k_3 R\sin\theta)$,
the explicit form (with $x\equiv k_3 R\sin\theta$) is
\begin{align}
I(\theta,\phi)
&= |\mathcal N g|^{2}\Big[
 n_{R+}\,J^2_{m^{(+)}_{R}}(x)+n_{R-}\,J^2_{m^{(-)}_{R}}(x)
 +2\,|\Gamma_{R}|\,J_{m^{(+)}_{R}}(x)J_{m^{(-)}_{R}}(x)\,
 \cos\big( \Delta m_{R}\,\phi+\psi_R \big)
\Big]\nonumber\\
&\quad+|\mathcal N g|^{2}\Big[
 n_{L+}\,J^2_{m^{(+)}_{L}}(x)+n_{L-}\,J^2_{m^{(-)}_{L}}(x)
 +2\,|\Gamma_{L}|\,J_{m^{(+)}_{L}}(x)J_{m^{(-)}_{L}}(x)\,
 \cos\big( \Delta m_{L}\,\phi+\psi_L \big)
\Big].
\label{eq:qI_explicit}
\end{align}
Here
\begin{equation}
\begin{aligned}
\Delta m_{R}&=m^{(-)}_{R}-m^{(+)}_{R}=2(M-N),\\
\Delta m_{L}&=m^{(-)}_{L}-m^{(+)}_{L}=2(M-N),
\end{aligned}
\end{equation}
so the interference fringes have azimuthal frequency $2|M-N|$. When $|M-N|=2$ the on-axis emission ($\theta\to 0$) is bright.

Under the bright-mode Hamiltonian, the internal conversion into the CW/CCW bright
supermodes is $\eta_{\rm DFG}^{(\pm)}=\sin^{2}(g_{\rm eff}^{(\pm)}T)$, and each radiated channel inherits the corresponding
fraction via its coupling weight, e.g.
$n_{R+}=\frac{|g^{(+)}_{R}|^{2}}{(g_{\rm eff}^{(+)})^{2}}\eta_{\rm DFG}^{(+)}\,n_{\rm in}$, etc.
The collected photon flux follows by integrating \eqref{eq:qI_explicit} over the NA cone,
giving the spatial factor $\eta_{\rm spatial}$ in the total efficiency
$\eta_{\rm tot}=\eta_{\rm ZPL}\,\eta_{\rm DFG}\,\eta_{\rm spatial}$.

\subsection*{C. Paraxial approximation details}
For completeness we summarise the paraxial-field formulas referenced above. Under this approximation, the positive-frequency near-field operator for each dipole at position $\phi_n$ is given by:
\begin{equation}
\hat{\mathbf E}^{(+)}_{i,n} = \mathcal{N}_i\, \hat a_{3,i}\, e^{i Q_i \phi_n}\, e^{-i \omega_3 t}\, |\sigma_i\rangle,
\label{eq:dipole_nf}
\end{equation}
where $\hat a_i$ is the annihilation operator for mode $i$, $Q_i$ is its associated angular momentum, $\sigma_i \in \{R, L\}$ denotes its circular polarization state, and $\mathcal{N}_i$ absorbs mode-dependent constants including the dipole amplitude and field normalization.

In the paraxial approximation and in the high-density limit \(N \rightarrow \infty\), the radiated field from the virtual Bragg grating can be treated as a continuous ring of azimuthally phased dipoles. The total far-field positive-frequency operator for each mode $i$ is then given by:

\begin{equation}
\hat{\mathbf E}^{(+)}_{\text{ff},i}
= \mathcal N_i\; g(\theta,r)\;
S_{Q_i}(\theta,\phi)\;\;\hat a_{3,i}\;e^{-i\omega_3 t} |\sigma_i\rangle,
\label{eq:ff_single_mode}
\end{equation}

where
\begin{equation}
g(\theta,r)=\frac{k_3^2}{4\pi\varepsilon_0}\,\frac{e^{ik_3 r}}{r}
L_z,
\qquad
S_{Q_i}(\theta,\phi)=(-i)^{Q_i}e^{iQ_i\phi}J_{Q_i}(k_3 R \sin\theta).
\end{equation}
Here $k_3 = \omega_3 / c$, $L_z$ is the effective film thickness, and $J_{Q_i}$ is the Bessel function of the first kind.

The total field $\hat{\mathbf E}^{(+)}_{\mathrm{ff}}=\sum_{i} \hat{\mathbf E}^{(+)}_{\text{ff},i}$ is the coherent sum over all sites. With the vectors $\mathbf v_{R/L}$ and matrices $\Gamma_{R/L}$ defined above, the intensity simplifies to

\begin{equation}
I(\theta,\phi)=\mathbf v_R^{\dagger} \Gamma_R \mathbf v_R+\mathbf v_L^{\dagger} \Gamma_L \mathbf v_L.
\label{eq:ff_intensity}
\end{equation}

In this notation the diagonal entries of $\Gamma_R$ ($\Gamma_L$) give the populations $n_{R\pm}$ ($n_{L\pm}$), while the off-diagonal entries carry the coherences $\Gamma_{R}=|\Gamma_{R}|e^{i\psi_R}$ and $\Gamma_{L}=|\Gamma_{L}|e^{i\psi_L}$. Writing $n_{R\pm}=\langle\hat a_{3,R\pm}^{\dagger}\hat a_{3,R\pm}\rangle$ and $n_{L\pm}=\langle\hat a_{3,L\pm}^{\dagger}\hat a_{3,L\pm}\rangle$ is convenient for the expressions below.
$n'_{3,+}=\langle\hat a_{3,+'}^\dagger\hat a_{3,+'}\rangle$, $n'_{3,-}=\langle\hat a_{3,-'}^\dagger\hat a_{3,-'}\rangle$,
and coherences $\Gamma_{-+}=|\Gamma_{-+}|e^{i\psi_R}$,
$\Gamma_{-'\!+\!'}=|\Gamma_{-'\!+\!'}|e^{i\psi_L}$,

\section*{III. Pump implementation}
\phantomsection\label{supp:secIII}

The programmable drive realizes the Fourier harmonics that shape the far field in Sec.~II, implemented with a spatial-light modulator (SLM) platform. The pump must (i) carry azimuthal phase $e^{im\phi}$ with $m=\ell N$ to satisfy the selection rule and (ii) overlap the WGM envelope at $r_0\simeq R$, so its tangential bandwidth is the central hardware constraint.

\subsection*{A. Bandwidth requirements and light-cone constraints}
The azimuthal phase demands tangential spatial frequency $k_\theta=m/r_0$ and numerical aperture
\[
  \mathrm{NA}_{\mathrm{req}} = \frac{m\lambda_p}{2\pi r_0} \simeq n_{\mathrm{eff}}\frac{\lambda_p}{\lambda_{\mathrm{res}}},
\]
which depends primarily on material index and the wavelength ratio rather than the disk size because both $m$ and $r_0$ scale with $R$. Angular momentum conservation fixes $m=\ell N$, and the output harmonics are $m'=\ell N\pm M\pm 2$; the $\Delta m=\pm2$ cases include an $m'=0$ on-axis component that feeds efficient fiber collection. Practical delivery is set by the smaller of the optical pupil cutoff $k_{\mathrm{opt}}=k_0\,\mathrm{NA}$ and the SLM sampling cutoff $k_{\mathrm{SLM}}=\pi/(pD)$, giving $k_c=\min(k_{\mathrm{opt}},k_{\mathrm{SLM}})$. For the representative LCOS setup detailed in Sec.~III.B (12.5\,$\mu$m pitch, $4f$ relay, $\mathrm{NA}=0.95$), the system operates near the objective limit, and the remaining gap to $\mathrm{NA}_{\mathrm{req}}\approx2.8$ sets the pump-efficiency penalty.

\begin{figure}[t]
  \centering
  \includegraphics[width=0.9\linewidth]{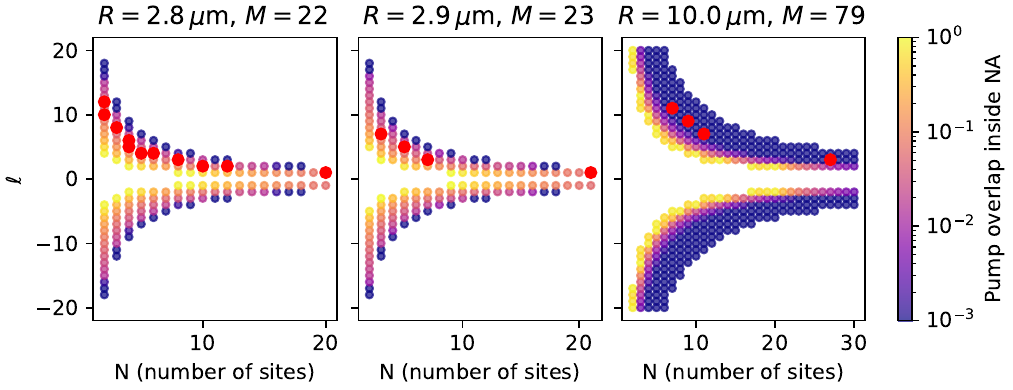}
  \caption{Feasibility of selection-rule harmonics in $(N,\ell)$ space for the present SLM/optic ($p=12.5\,\mu\mathrm{m}$, $f=1.5$\,mm, $P_{\max}=6$\,mm, $\lambda_{\mathrm{res}}=736$\,nm, $\mathrm{NA}=0.95$, $n_{\mathrm{eff}}=1.7$). Points satisfy $|m'|/R<k_0\mathrm{NA}$ for $m'=\ell N\pm M\pm2$. Color shows the pump harmonic’s $|J_m|^2$ fraction inside NA; $\Delta m=\pm2$ cases (with on-axis $m'=0$ components) are highlighted.}
  \label{fig:feasibility-map}
\end{figure}

\noindent Figure~\ref{fig:feasibility-map} visualises the allowed $(N,\ell)$ pairs and the pump-efficiency hit imposed by the NA limit. The red points mark the $\Delta m=\pm2$ channels that include the on-axis $m'=0$ harmonic needed for fiber collection, while the color bar reports the fraction of the pump harmonic that falls inside the objective NA. For our $N=23$ choice, the map highlights that only $\sim25\%$ of the Bessel spectrum is delivered, consistent with the \(\eta_{\mathrm{pump}}\approx0.25\) factor used below.

\subsection*{B. Optical layout and calibration}
A possible experimental implementation could use a reflective LCOS phase SLM operating at \(1.6\,\mu\mathrm{m}\) (pixel pitch \(12.5\,\mu\mathrm{m}\)) which encodes a double-phase hologram and is relayed onto the device with a \(4f\) system followed by a high-NA objective (\(\mathrm{NA}=0.95\)). Choosing a demagnification of \(40\times\) maps each SLM pixel to \(\approx0.31\,\mu\mathrm{m}\) in the sample plane. The focused pump waist is \(w_0\simeq0.6\,\mu\mathrm{m}\).

We note that experimentally creating 23 spatially separated intensity peaks would demand tangential bandwidth \(\mathrm{NA}_{\mathrm{lobes}}\approx N\lambda_p/(2\pi r_0)\approx 2.8\), exceeding the objective limit \(\mathrm{NA}=0.95\). However, in practice 23 discrete sites are not necessary, and an experimental configuration in which pump imprints only the required helical phase structure \(e^{i23\phi}\) around a continuous ring at radius \(r_0\approx R\) is sufficient and supplies the momentum-matching condition without requiring discrete sites. The resulting pump coupling efficiency is reduced by a factor of \(\sim4\) relative to an ideal Bessel beam (Sec.~III.E), but this penalty can be compensated by proportionally higher pump power.  In contrast, the converted single photons couple efficiently to a fiber-coupled detector because the \(m'=0\) output harmonic is on-axis and well-captured by the objective NA. At tight focusing, spin-orbit coupling of light introduces OAM-SAM mixing~\cite{bliokh2015spinorbit}; this is implicitly included in the NA-limited model and further motivates the smooth-annulus strategy when \(m\) is high. (See Fig.~\ref{fig:polarization-story} for a compact visualization of the focal and far-field polarization geometry.)

The diffraction-limited Gaussian waist at \(\lambda_p=1623\) nm and \(\mathrm{NA}=0.95\) is \(w_0 \gtrsim 0.52\,\mu\)m, so experimental estimates use \(w_0\approx0.6\,\mu\)m. The idealised FDTD cases with \(120\,\mathrm{nm}\) site waists serve only to isolate selection-rule physics; the far-field depends on the azimuthal order \(m\), and the experimentally feasible smooth-annulus pump (larger waist) produces the same \(m=23\) harmonics with the pump-efficiency penalty quantified in Sec.~III.E. Programmability follows from updating the hologram; present LCOS devices refresh at \(60\) to \(120\) Hz so the virtual Bragg grating can be reconfigured on sub-second timescales. The impact of imperfect phase control was assessed by adding random offsets \(\delta\phi_n\) to the pump harmonics and recomputing Eq.\,\eqref{eq:qI_explicit}; a phase standard deviation of \(5^{\circ}\) reduces the far-field peak by less than \(3\%\). Thermally stabilised SLMs combined with slow feedback are specified to hold the pixel phase within \(<1^{\circ}\), indicating that the azimuthal profile assumed in the simulations is compatible with current experimental practice. In situ calibration can be performed by imaging the reflected pump pattern and iteratively adjusting the hologram until the desired Fourier spectrum is obtained, while the converted telecom output is monitored to lock the overall phase.

\subsection*{C. Design examples}
Bandwidth limits can be addressed with three complementary approaches. (1) A smooth annular pump that encodes the full helical phase maximises WGM overlap when \(\mathrm{NA}_{\mathrm{cap}}\ge\mathrm{NA}_{\mathrm{req}}\). (2) Offloading helical charge with a q-plate supplies \(\Delta m=2\sigma q\), leaving the SLM to encode \(m_{\mathrm{SLM}}=m-2\sigma q\) with reduced \(\mathrm{NA}_{\mathrm{phase}}=|m_{\mathrm{SLM}}|\lambda_p/(2\pi r_0)\); the minimum charge is \(q_{\min}=\lceil(|m|-k_0 r_0\,\mathrm{NA}_{\mathrm{cap}})/2\rceil_+\). (3) Optional discrete \(N\)-lobe modulation introduces a lobe-spacing burden \(\mathrm{NA}_{\mathrm{lobes}}\approx N\lambda_p/(2\pi r_0)\) and is only needed for site-resolved addressing.

For the main-text configuration (\(M=21\), \(\ell=1\), \(N=23\), \(r_0\approx1.3\,\mu\mathrm{m}\)), the pump order is \(m=23\) with \(\mathrm{NA}_{\mathrm{req}}\approx2.8\). Using a \(q=5\) plate leaves \(m_{\mathrm{SLM}}=13\), so truncation at \(\mathrm{NA}=0.95\) yields \(\eta_{\mathrm{pump}}\approx0.25\), i.e., a \(\sim4\times\) power penalty that is acceptable because the converted photons remain on-axis. Fewer-site designs (e.g., \(N=2\) or \(N=3\)) follow the same trade-off: q-plates with \(q\lesssim5\) tame the helical ramp, while smooth annuli avoid additional lobe-spacing bandwidth.

\begin{figure}[t]
  \centering
  \includegraphics[width=\linewidth]{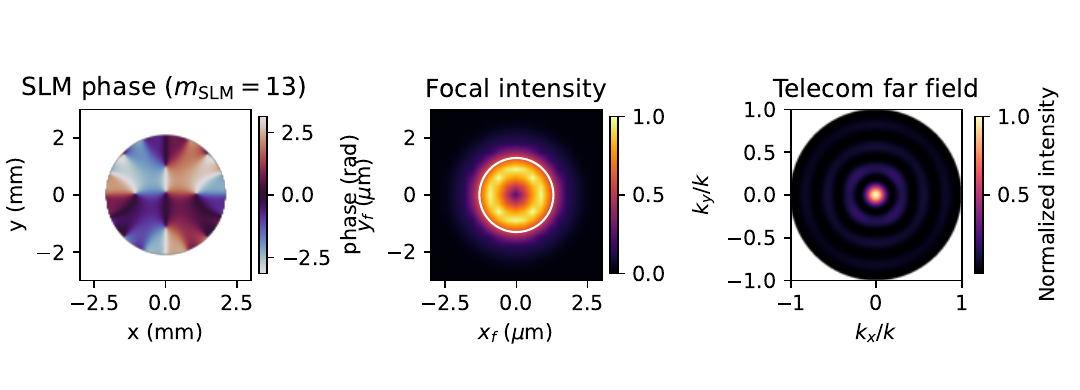}
  \caption{Design example for the main-text configuration (\(N=23\), \(\ell=1\), \(q=5\), \(M=21\)). Left: phase-only SLM pattern encoding the residual \(m_{\mathrm{SLM}}=13\) after q-plate offload. Middle: resulting focal-plane intensity (linear scale) forming the helical ring at \(r_0\simeq1.3\,\mu\)m; the white circle marks the disk rim. The NA-limited pupil admits \(\approx28\%\) of the pump power for this design (consistent with the \(\eta_{\mathrm{pump}}\approx0.25\) estimate in Sec.~III.C). Right: telecom far-field emission showing the on-axis \(m'=0\) component with weak \(|m'|=4\) sidelobes.}
  \label{fig:design-triptych}
\end{figure}

Figure~\ref{fig:design-triptych} illustrates this configuration: the SLM encodes the residual \(m_{\mathrm{SLM}}=13\) phase, the focus forms the required helical ring at the disk rim, and the telecom far field is dominated by the on-axis \(m'=0\) component with weak \(|m'|=4\) sidelobes.

\subsection*{D. Addressable array size for N\,$\times$\,1 switching}
A linear phase ramp (blaze) added to the SLM hologram steers the focal spot across the sample. The maximum usable span is set by the objective acceptance, \(\Delta x_{\max} \simeq 2 f\,\mathrm{NA}\); with \(f=1.5\) mm and \(\mathrm{NA}=0.95\), \(\Delta x_{\max}\approx 2.85\) mm. A square grid with \(10\,\mu\)m pitch therefore supports
\begin{equation}
N_{\mathrm{addr}} \approx \left(\frac{\Delta x_{\max}}{10\,\mu\mathrm{m}}\right)^{2} \approx 8\times10^{4}
\label{eq:naddr}
\end{equation}
addressable sites without moving hardware. SLM refresh (60 to 120 Hz) reconfigures the steering phase on sub-second timescales; the practical limit is set by field uniformity across the span rather than the hologram itself.

\subsection*{E. Pump-power scaling}
The overlap integrals extracted from the DFG model give an ideal coupling rate \(G_{\mathrm{ideal}}\approx 2\pi\times 60\,\mathrm{MHz}\) for a perfect Bessel beam matched to the WGM. Hardware bandwidth limitations reduce the delivered pump coupling efficiency by a factor of \(\eta_{\mathrm{pump}}\approx 0.25\), requiring the effective coupling rate \(G_{\mathrm{eff}}=\sqrt{\eta_{\mathrm{pump}}}\,G_{\mathrm{ideal}}\approx 2\pi\times 30\,\mathrm{MHz}\) or equivalently \(\sim4\times\) higher pump power to reach the same nonlinear conversion efficiency. Reaching \(\eta_{\mathrm{DFG}}=0.5\) therefore requires a total pump power of \(P_{\mathrm{tot}}\approx 180\) mW distributed across the azimuthal Fourier components. With the pump waist \(w_0=0.6\,\mu\mathrm{m}\) and \(N=23\) harmonics, the local intensity per site is
\begin{equation}
I_{\mathrm{site}}\approx\frac{(180\,\mathrm{mW})/23}{\pi w_0^{2}}\simeq 7\times10^{5}\,\mathrm{W/cm^2},
\label{eq:I_site}
\end{equation}
still well below CW damage thresholds (\(\gtrsim10^{7}\,\mathrm{W/cm^2}\)) reported for thin-film LiNbO\(_3\). Allowing for \(\sim30\%\) SLM reflectivity and \(90\%\) transmission through the relay optics requires an incident laser power of \(\sim700\) mW, readily available from commercial telecom fiber lasers.

Evaluating Eqs.\,(13) to (15) with the numerically computed WGM profiles gives a linear relation between the peak pump field at the disk rim and the bright-mode coupling rate,
\begin{equation}
G_{\mathrm{eff}} = \Lambda\,\frac{\sqrt{N\,P_{\mathrm{tot}}}}{w_0},
\label{eq:Geff_scaling}
\end{equation}
where \(P_{\mathrm{tot}}\) is the total optical power delivered by the SLM pattern, \(w_0\) is the pump waist on the rim, \(N\) is the number of azimuthal harmonics, and the constant \(\Lambda = 2.3\times10^{2}\,\mathrm{s^{-1}}\sqrt{\mathrm{m^2/W}}\) is obtained from the overlap integrals. For the parameters used in Fig.\,2 (\(N=23\), \(w_0=0.6\,\mu\mathrm{m}\)) the expression reduces to
\begin{equation}
G_{\mathrm{eff}} \approx 2\pi\times 60\,\mathrm{MHz}
\left(\frac{P_{\mathrm{tot}}}{45\,\mathrm{mW}}\right)^{1/2}
\left(\frac{23}{N}\right)^{1/2}
\left(\frac{0.6\,\mu\mathrm{m}}{w_0}\right).
\label{eq:Geff_numeric}
\end{equation}
Setting \(G_{\mathrm{eff}} = 2\pi\times60\) MHz therefore requires \(P_{\mathrm{tot}}\approx 45\) mW. Distributing this power evenly among the harmonics gives \(P_{\mathrm{site}} = P_{\mathrm{tot}}/N \approx 2\) mW per harmonic, which leads to a local intensity \(I_{\mathrm{site}} = 2P_{\mathrm{site}}/(\pi w_0^2)\approx 1.8\times10^{5}\,\mathrm{W/cm^2}\). The scaling also shows that larger waists or a higher value of \(N\) increase the required pump power only as the square root, while time-domain control remains unaffected because the hologram update rate is set by the SLM electronics.

Long nanosecond pump pulses offer a quasi-CW route to high peak power with low average power, compatible with SLM refresh~\cite{mcguinness2010pcfqfc,reddy2013pplnqfc}. Such formats can supply the \(\sim 100\) mW-level total power estimated here without exceeding damage or thermal limits.

\subsection*{F. Reconfiguration speed and bandwidth}
\textbf{SLM refresh.} The reflective LCOS devices discussed above refresh at 60 to 120\,Hz, setting a sub-second ceiling on reconfiguration. Faster updates are available with segmented electro-optic pump waveguides, which support kHz tuning at the cost of additional fabrication complexity.

\textbf{Spectral bandwidth.} The conversion bandwidth follows the cavity response, with efficiency
\begin{equation}
\eta_{\mathrm{DFG}}(\Delta)=\frac{\eta_0}{1+(2\Delta/\kappa)^2}
\label{eq:eta_detuning}
\end{equation}
where \(\kappa/2\pi\approx 400\,\mathrm{GHz}\) for \(\mathcal Q\approx 10^3\) at \(736\,\mathrm{nm}\). Thus a detuning window of about \(100\,\mathrm{GHz}\) retains \(\eta_{\mathrm{DFG}}>0.5\).

Thermal or electro-optic tuning of the WGM resonance can accommodate slow drift without requiring pump-pattern updates.

\textbf{Efficiency breakdown.} The total end-to-end conversion efficiency decomposes into three factors: $\eta_{\mathrm{tot}} = \eta_{\mathrm{ZPL}} \cdot \eta_{\mathrm{DFG}} \cdot \eta_{\mathrm{spatial}}$, summarized in Table~\ref{tab:efficiency}. For the present design with $Q \approx 10^3$, the nonlinear conversion remains the bottleneck ($\eta_{\mathrm{DFG}} \sim 10^{-8}$); increasing $Q$ to $2 \times 10^6$ would raise $\eta_{\mathrm{DFG}}$ to $0.1$, yielding $\eta_{\mathrm{tot}} \approx 7\%$.

\begin{table}[h]
\centering
\caption{Efficiency breakdown for the virtual Bragg grating interface with $N=23$, $M=21$, and $Q=10^3$. Values are computed from FDTD simulations (Sec.~V) and analytical expressions (Secs.~I and II). The bottleneck is $\eta_{\mathrm{DFG}}$, which scales as $\sin^2(g_{\mathrm{eff}}T) \propto Q$.}
\begin{tabular}{llll}
\hline\hline
Component & Symbol & Value & Notes \\
\hline
Zero-phonon line coupling & $\eta_{\mathrm{ZPL}}$ & $0.99$ & SiV Debye-Waller factor \\
DFG conversion (current $Q$) & $\eta_{\mathrm{DFG}}$ & $2.2 \times 10^{-8}$ & $Q=10^3$, $g_{\mathrm{eff}}=2\pi \times 60$\,MHz \\
Spatial coupling efficiency & $\eta_{\mathrm{spatial}}$ & $0.70$ & Far-field overlap with NA (Fig.~2) \\
\hline
\textbf{Total (current $Q$)} & $\eta_{\mathrm{tot}}$ & $1.5 \times 10^{-8}$ & Product of above \\
\hline
DFG conversion (high $Q$) & $\eta_{\mathrm{DFG}}$ & $0.10$ & $Q=2.2 \times 10^6$ \\
\textbf{Total (high $Q$)} & $\eta_{\mathrm{tot}}$ & $0.069$ & With improved cavity \\
\hline\hline
\end{tabular}
\label{tab:efficiency}
\end{table}

\section*{IV. Noise and decoherence estimates}
\phantomsection\label{supp:secIV}

The pump powers and spatial profiles established above bound the dominant parasitic channels, quantified below.

\textbf{Raman scattering.} Lithium niobate exhibits a Raman gain coefficient \(g_R \approx 5\,\mathrm{cm/GW}\) for the dominant optical phonon modes at telecom wavelengths~\cite{dhar1985stimulated}. Using the SLM-delivered intensity \(I_{\mathrm{site}}\approx 1.8\times10^{5}\,\mathrm{W/cm^2}\) and an interaction length equal to the nonlinear film thickness \(L_z = 0.28\,\mu\mathrm{m}\), the spontaneous Raman conversion fraction is
\[
f_R \approx g_R\,I_{\mathrm{site}}\,L_z \approx 5\times10^{-11}\,\frac{\mathrm{m}}{\mathrm{W}} \times 1.8\times10^{9}\,\frac{\mathrm{W}}{\mathrm{m^2}} \times 2.8\times10^{-7}\,\mathrm{m} \simeq 3\times10^{-8}.
\]
Thus fewer than \(10^{-7}\) Raman photons are generated per converted telecom photon, even before applying spectral filtering. The residual Raman contribution can therefore be neglected to first order in the conversion efficiency budget.

\textbf{Pump leakage.} The pump and signal wavelengths are separated by \( \sim 900\,\mathrm{nm}\), enabling aggressive spectral filtering. Cascaded volume-Bragg and etalon filters routinely provide \(>120\,\mathrm{dB}\) pump extinction in quantum frequency conversion experiments~\cite{pelc2011long}. Applying this value to our \(45\,\mathrm{mW}\) pump leaves \(<4\times10^{-15}\,\mathrm{W}\) in the signal band, or \(3.7\times10^{4}\) photons~s\(^{-1}\) at \(1623\,\mathrm{nm}\). This is comparable to, or below, the dark-count rate of modern SNSPDs, so pump leakage does not limit single-photon fidelity provided standard telecom filtering is used.

\textbf{Thermal occupation and phonon sidebands.} The spin-photon interface is operated at cryogenic temperature (\(<5\,\mathrm{K}\)) to address the SiV zero-phonon line~\cite{sipahigil2016integrated}. At this temperature the thermal occupancy of the \(736\,\mathrm{nm}\) WGM satisfies \(n_{\mathrm{th}} \approx \exp[-\hbar\omega/(k_B T)] < 10^{-80}\), so blackbody photons are negligible. Residual phonon sidebands are suppressed both by the cavity linewidth and by the narrow intrinsic linewidth of the SiV center, which has demonstrated Debye-Waller factors \(>70\%\) and phonon-limited linewidths below \(100\,\mathrm{MHz}\) under similar conditions~\cite{hepp2014sivstructure}. Consequently the main decoherence channel remains pump-noise-induced scattering, which, as shown above, is strongly suppressed by the choice of material and filtering.

\section*{V. Numerical verification}
\phantomsection\label{supp:secV}

\subsection*{A. Simulation methods and isotropic surrogate}

To verify the analytical predictions for the bright-mode coupling [Eq.~\eqref{eq:H_bright_single}] and the far-field structure [Eq.~\eqref{eq:qI_explicit}], we perform finite-
difference time-domain (FDTD) simulations in Lumerical. The present release of the software implements the second-order susceptibility only for isotropic media. We therefore emulate the LiNbO$_3$ response with a fictitious scalar nonlinearity $\chi_{\mathrm{iso}}$ chosen to reproduce the root-mean-square magnitude of the anisotropic coefficients, $\chi_{\mathrm{iso}} = \sqrt{(|d_{R}|^{2}+|d_{L}|^{2})/2}$. In practice we set $\chi_{11} = \chi_{22} = 4.5 \times 10^{-11}$\,m/V (all other tensor elements zero), which matches the RMS value of the rotated tensor at the operating wavelength.

The mesh size is set to 10\,nm inside the cavity and 50\,nm in the surrounding air with a PML boundary condition. The simulation domain is performed in a box  $10\,\mu\text{m} \times 10\,\mu\text{m} \times 6\,\mu\text{m}$ with a PML boundary condition on the boundaries. Simulations were run on our university's computing cluster equipped with \textit{AMD EPYC 7763} processors, with each run averaging 4\,hours. Dispersive material properties are used for $\text{LiNbO}_{3}$, assuming an ordinary refractive index in the in-plane directions and an extraordinary index along the axial direction.

The pump field is modeled as \(N=23\) discrete Gaussian beams positioned at \(\phi_n = 2\pi n/N\), each carrying the appropriate phase \(\exp(i\ell\phi_n)\) to generate the virtual Bragg grating.  The simulated conversion amplitudes are subsequently re-scaled phenomenologically by the analytic factors $d_{R}/\chi_{\mathrm{iso}}$ and $d_{L}/\chi_{\mathrm{iso}}$ when comparing against Eqs.~\eqref{eq:H_bright} to \eqref{eq:split_ccw}. This preserves the correct anisotropic weighting while keeping the numerics within the available FDTD feature set.

Under these assumptions, the nonlinear source reduces
to an array of in-plane linear dipoles oriented at an angle
of $-45^{\circ}$ with respect to the $\hat x$-axis. This corresponds to
the polarization state: $\ket{-45^{\circ}} = \tfrac{1}{\sqrt{2}}(\hat x- \hat y)$.

The total radiation field is the superposition of the
fields resulting from both CW and CCW propagating
components:
\begin{equation}
\vec E_{\mathrm{CW}} = -iN g
\Big[
(-i)^{Q_{+}} e^{iQ_{+}\phi} J_{Q_{+}}
(k R \sin \theta) + (-i)^{Q'_{+}} e^{iQ'_{+}\phi} J_{Q'_{+}}
(k R \sin \theta)
\Big]
\ket{-45^{\circ}},
\label{eq:E_cw}
\end{equation}
\begin{equation}
\vec E_{\mathrm{CCW}} = -iN g
\Big[
(-i)^{Q_{-}} e^{iQ_{-}\phi} J_{Q_{-}}(k R \sin \theta) + (-i)^{Q'_{-}} e^{iQ'_{-}\phi} J_{Q'_{-}}
(k R \sin \theta)
\Big]
\ket{-45^{\circ}}.
\label{eq:E_ccw}
\end{equation}
The total electric field is given by: $\vec E = \vec E_{\mathrm{CW}}+ \vec E_{\mathrm{CCW}}$. The corresponding far-field intensity is:
\begin{equation}
\begin{aligned}
I(x, \phi) = 2A \big[
&J^{2}_{Q_{+}}(x) + J^{2}_{Q'_{+}}(x) + J^{2}_{Q_{-}}(x) + J^{2}_{Q'_{-}}(x)\\
&+ 2J_{Q_{+}}(x)J_{Q'_{+}}(x) \cos [(Q_{+} -Q'_{+})\phi] + 2J_{Q_{+}}(x)J_{Q_{-}}(x) \cos [(Q_{+} -Q_{-})\phi]\\
&+ 2J_{Q_{+}}(x)J_{Q'_{-}}(x) \cos [(Q_{+} -Q'_{-})\phi] + 2J_{Q'_{+}}(x)J_{Q_{-}}(x) \cos [(Q'_{+} -Q_{-})\phi]\\
&+ 2J_{Q'_{+}}(x)J_{Q'_{-}}(x) \cos [(Q'_{+} -Q'_{-})\phi]
\big],
\end{aligned}
\label{eq:I_full}
\end{equation}
where $A$ is a constant and $x = k R \sin \theta$.
Among the five possible interference terms, only two
contribute significantly for arbitrary combinations of az-
imuthal mode numbers and virtual grating indices. The
cross-terms involving Bessel functions of different orders vanish due to their orthogonality. Therefore, the
terms containing products such as $J_{Q_{+}}J_{Q'_{+}}$, $J_{Q_{+}}J_{Q_{-}}$,
and $J_{Q'_{+}}J_{Q'_{-}}$
are negligible. By using the relations $Q_{-} =
- Q'_{+}$ and $Q'_{-} = - Q_{+}$, the intensity simplifies to:
\begin{equation}
I(x, \phi) = 2A
\big[
J^{2}_{Q_{+}}(x) + J^{2}_{Q'_{+}}(x) + J^{2}_{Q_{-}}(x) + J^{2}_{Q'_{-}}(x)
+ 2J^{2}_{Q_{+}}(x) \cos (2Q_{+}\phi) + 2J^{2}_{Q_{-}}(x) \cos (2Q_{-}\phi)
\big].
\label{eq:I_simplified}
\end{equation}
For $N > M$, we have $|Q_{+}| < |Q_{-}|$. Thus, the interference pattern in low NA, is primarily determined by
the term $J^{2}_{Q_{+}}(x) \cos(2Q_{+}\phi)$, which has a periodicity of
$\pi/Q_{+}$. Moreover, for $N < M$, the dominant contribu-
tion comes from $J^{2}_{Q_{-}}(x) \cos(2Q_{-}\phi)$.

\subsection*{B. Device parameters and validation results}

\paragraph*{Simulation setup.}
We benchmark the analytical expressions with finite-difference time-domain simulations in Lumerical. The geometry reproduces the device discussed in the main text: a diamond ring of radius $R=1.6\,\mu$m with width $w=200$\,nm and thickness $t_1=100$\,nm, clad by a $t_2=280$\,nm z-cut LiNbO$_3$ ring. The structured pump is represented by $N$ Gaussian beams (waist $a=120$\,nm) arranged with the same azimuthal periodicity used in Sec.~I, each carrying identical power at a wavelength of $1623$\,nm and peak amplitude $10^{9}$\,V/m. Table~\ref{tab:device_params} summarizes the key device and simulation parameters.

\begin{table}[h]
\centering
\caption{Device and simulation parameters for the hybrid diamond-LiNbO$_3$ WGM resonator and structured pump configuration used in Figs.~1 to 3 of the main text.}
\begin{tabular}{ll}
\hline\hline
Parameter & Value \\
\hline
Resonator radius, $R$ & $1.6\,\mu$m \\
Resonator width, $w$ & $200$\,nm \\
Diamond layer thickness, $t_1$ & $100$\,nm \\
LiNbO$_3$ layer thickness, $t_2$ & $280$\,nm \\
Pump wavelength, $\lambda_2$ & $1623$\,nm \\
Total pump power, $P_{\mathrm{tot}}$ & $45$\,mW \\
Pump beam waist (per site), $a$ & $120$\,nm \\
Fiber mode waist, $w_0$ & $0.6\,\mu$m \\
Azimuthal sampling, $N$ & $23$ (Figs.~1c, 3); varied (Fig.~2) \\
Dominant WGM mode, $(M,p,q)$ & $(21,1,2)$ at $736$\,nm \\
Telecom output wavelength, $\lambda_3$ & $1347$\,nm \\
Quality factor, $Q$ & $\sim 10^3$ \\
\hline\hline
\end{tabular}
\label{tab:device_params}
\end{table}

\begin{table}[h]
\centering
\caption{Material properties for diamond and thin-film LiNbO$_3$ at the operating wavelengths. Refractive indices are used for mode calculations and phase-matching conditions; $\chi^{(2)}$ tensor elements determine RHCP/LHCP coupling coefficients $d_{R,L}$.}
\begin{tabular}{llll}
\hline\hline
Material & Property & Value & Wavelength \\
\hline
Diamond & Refractive index, $n_{\mathrm{dia}}$ & $2.41$~\cite{zaitsev2001optical} & $736$\,nm \\
 & Refractive index, $n_{\mathrm{dia}}$ & $2.38$~\cite{zaitsev2001optical} & $1347$\,nm \\
 & Refractive index, $n_{\mathrm{dia}}$ & $2.38$~\cite{zaitsev2001optical} & $1623$\,nm \\
\hline
LiNbO$_3$ & Refractive index (ordinary), $n_o$ & $2.23$~\cite{gayer2008temperature} & $736$\,nm \\
(z-cut) & Refractive index (ordinary), $n_o$ & $2.14$~\cite{gayer2008temperature} & $1347$\,nm \\
 & Refractive index (ordinary), $n_o$ & $2.13$~\cite{gayer2008temperature} & $1623$\,nm \\
 & Nonlinear coefficient, $d_{22}$ & $2.1 \times 10^{-12}$\,m/V~\cite{nikogosyan2005nonlinear} & n/a \\
 & Nonlinear coefficient, $d_{31}$ & $-4.3 \times 10^{-12}$\,m/V~\cite{nikogosyan2005nonlinear} & n/a \\
 & Effective (isotropic), $\chi_{\mathrm{iso}}$ & $4.5 \times 10^{-11}$\,m/V$^\dagger$ & FDTD surrogate \\
\hline
Effective & WGM index, $n_{\mathrm{eff}}$ & $1.54$ & $736$\,nm (hybrid mode) \\
\hline\hline
\end{tabular}
\vspace{2pt}
\begin{flushleft}
\footnotesize $^\dagger$Calibrated to reproduce anisotropic LiNbO$_3$ response within the isotropic FDTD surrogate used in Sec.~V.
\end{flushleft}
\label{tab:materials}
\end{table}

\paragraph*{Spectral features.}
Panel (b) of Fig.~1 in the main text reports the converted spectrum obtained from this model. Several neighboring WGMs around 736\,nm participate, but the resonance at 736\,nm dominates, producing a strong converted line at 1347\,nm ($1/739\,\mathrm{nm}+1/1623\,\mathrm{nm}\rightarrow1/1347\,\mathrm{nm}$) with a simulated conversion efficiency of $17\%$ for the classical drive levels considered here.

\paragraph*{Far-field structure.}
The simulated far-field patterns for virtual grating orders $N=22$, $23$, and $25$ (Fig.~2, main text) track the analytic predictions: $M-N=-1$ yields a two-lobed pattern, $M-N=-2$ gives a bright on-axis spot, and $M-N=-4$ produces the expected fourfold symmetry. Deviations at large $\theta$ arise from the finite numerical aperture and from the residual anisotropy neglected in the isotropic surrogate used by the solver. Table~\ref{tab:farfield_patterns} summarizes the programmable far-field characteristics for different pump configurations.

\begin{table}[h]
\centering
\caption{Far-field emission patterns as a function of pump azimuthal sampling $N$ with fixed $M=21$ and $\ell=1$. The selection rule $M = \ell N \pm 2$ determines the azimuthal harmonic order $\Delta m = 2|M - \ell N|$, which controls the number of lobes and on-axis brightness. Spatial coupling efficiency $\eta_{\mathrm{spatial}} = 0.70$ is reported for $N=23$ in the main text.}
\begin{tabular}{cccl}
\hline\hline
$N$ & $M - \ell N$ & $\Delta m$ & Pattern description \\
\hline
19 & $+2$ & 4 & On-axis bright spot \\
20 & $+1$ & 2 & Two-lobe dipole pattern \\
21 & $0$ & 0 & Ring (no on-axis emission) \\
22 & $-1$ & 2 & Two-lobe (FDTD, Fig.~2) \\
23 & $-2$ & 4 & On-axis bright spot (FDTD, Fig.~2) \\
25 & $-4$ & 8 & Four-lobe cloverleaf (FDTD, Fig.~2) \\
27 & $-6$ & 12 & Six-lobe pattern \\
\hline\hline
\end{tabular}
\label{tab:farfield_patterns}
\end{table}

\paragraph*{In-coupling configuration.}
For the write-in process (Fig.~3, main text) we illuminate the cavity from $-z$ with a shaped fiber mode at 1350\,nm, polarized along $\hat{\boldsymbol{\rho}}$, while the pump impinges from $+z$ with $\hat{\boldsymbol{\phi}}$ polarization. Both patterns use $N=23$ lobes. The target WGM at 738\,nm emerges with azimuthal number $M=20$ rather than $M=21$ because the strong classical fields induce a small Kerr-like shift of the effective index via cascaded $\chi^{(2)}$ processes.

The present device operates with a modest cavity quality factor, \(Q \approx 10^{3}\). Together with the intrinsically weak single-photon field emitted by the SiV center, this necessitated an artificially large dipole amplitude in the classical simulations to visualise the nonlinear dynamics. A more practical route toward high probability conversion is to increase \(Q\) (or, equivalently, the coherent coupling rate \(g_{\mathrm{eff}}\)).

Evaluating the bright-mode expression \(\eta_{\mathrm{tot}}=\eta_{\mathrm{ZPL}}\eta_{\mathrm{DFG}}\eta_{\mathrm{spatial}}\) with the manuscript parameters (\(\lambda_1=736\,\mathrm{nm}\), \(Q=10^{3}\), \(g_{\mathrm{eff}}=2\pi\times60\,\mathrm{MHz}\), \(\eta_{\mathrm{ZPL}}=0.99\), \(\eta_{\mathrm{spatial}}=0.70\)) yields a cavity lifetime \(T = Q/\omega_1 \approx 3.9\times10^{-13}\,\mathrm{s}\). The interaction strength gives \(g_{\mathrm{eff}}T \approx 1.5\times10^{-4}\), so \(\eta_{\mathrm{DFG}}=\sin^{2}(g_{\mathrm{eff}}T)\approx2.2\times10^{-8}\) and the overall conversion probability is \(\eta_{\mathrm{tot}}\approx1.5\times10^{-8}\). Achieving \(\eta_{\mathrm{DFG}}=0.1\) with the same coupling would require \(Q\approx2.2\times10^{6}\) (or, alternatively, \(g_{\mathrm{eff}}\approx2\pi\times130\,\mathrm{GHz}\) at fixed \(Q\)). State-of-the-art thin-film lithium niobate microrings already exhibit quality factors above this threshold~\cite{zhang2017monolithic}, indicating that the requisite cavity performance is experimentally accessible.

These observations reinforce the interpretation of the programmable pump as a virtual Bragg grating: by engineering both its spatial harmonics and polarization, the device accesses selection rules that would be difficult to realize with static lithographic gratings, while retaining rapid reconfigurability.

\bibliographystyle{apsrev4-2}

\bibliographystyle{apsrev4-2}
\bibliography{Ref3}

\end{document}